\pdfoutput=1
\documentclass[aps,prd,reprint,amsmath,amssymb,nofootinbib,superscriptaddress,onecolumn,notitlepage]{revtex4-1}
\usepackage{custom}

\begin{document}

\title{On the QCD critical point, Lee-Yang edge singularities and Pad\'e resummations}

\author{G\"{o}k\c{c}e Ba\c{s}ar}
\email{gbasar@unc.edu}
\affiliation{Department of Physics and Astronomy, University of North Carolina, Chapel Hill, North Carolina 27599, USA}

\date{\today}

%\newpage

\begin{abstract}  
I analyze the trajectory of the Lee-Yang edge singularities of the QCD equation of state in the complex baryon chemical potential ($\mu_B$) 
plane for different values of the temperature by using the recent lattice results for the Taylor expansion coefficients up to eighth order in $\mu_B$ 
and various resummation techniques that blend in Pad\'e expansions and conformal maps. 
By extrapolating from this information, I estimate for the location of the QCD critical point, $T_c\approx 100$ MeV, $\mu_c\approx 580$ MeV. I also estimate the crossover slope at the critical point to be $\alpha_1\approx 9^\circ$ and further constrain the non-universal mapping parameters between the  three-dimensional Ising model and QCD equations of state. 
\end{abstract}

\maketitle

\section{Introduction}
\label{sec:intro}

Mapping the phase diagram of Quantum Chromodynamics (QCD) for different temperatures and baryon densities, probed by the baryon chemical potential, $\mu_B$,  is currently one of the major open problems in nuclear physics. 
State-of-the-art lattice simulations offer extensive information about the thermodynamic properties at vanishing chemical potential. For instance, it is well established that the transition between the baryonic phase and quark-gluon plasma is a smooth crossover \cite{Aoki:2006we}.  However, the applicability of lattice QCD at finite chemical potential is severely limited due to the fermion sign problem (for reviews see e.g.  Refs. \cite{Philipsen:2007aa,deForcrand:2009zkb,Ding:2015ona} ). A central question regarding the QCD phase diagram at finite temperature and chemical potential is whether there is a second order critical point, and a first-order transition curve emanating from it \cite{Bzdak:2019pkr}. From the experimental perspective, this question has been the main motivation behind the Beam Energy Scan program at the Relativistic Heavy Ion Collider as well as the future Compressed Baryonic Matter experiment at the Facility for Antiproton and Ion Research \cite{Almaalol:2022xwv}.

From the theoretical standpoint, the first principle computations of the QCD phase diagram at finite chemical potential requires tackling the fermion sign problem in some way. The two main approaches that bypass the sign problem are Taylor expanding around $\mu_B=0$ \cite{Allton:2002zi}, and performing lattice simulations at pure imaginary chemical potential and analytically continuing from it \cite{deForcrand:2002hgr,DElia:2002tig,Bellwied:2015rza,Ratti:2018ksb,Borsanyi:2021sxv}. In the former approach the equation of state is expanded in $\mu_B$\footnote{More precisely the expansion is in $\mu_B^2$ due to charge conjugation symmetry.} as a Taylor series whose coefficients are observables evaluated at $\mu_B=0$ which can be computed without a sign problem. At the same time, even in the absence of the fermion sign problem, the computation of the higher order Taylor coefficients on the lattice becomes progressively more difficult due to noise. The state-of-the-art computations go up to ${\cal O}({\mu_B^8})$ (for recent results see HotQCD \cite{Bollweg:2022rps,Bollweg:2022fqq} and Wuppertal-Budapest  \cite{Borsanyi:2020fev} Collaborations). Currently, based on the available data, the existence of a critical point for $\mu_B/T\lesssim 3$ is highly unlikely.

Even with access to a modest number Taylor coefficients, it is possible extract valuable information regarding the equation of state that goes beyond approximating it as a truncated Taylor expansion. The starting point is that, for any temperature, the equation of state possess singularities for \textit{complex} values of $\mu_B^2$, the Lee-Yang (LY) edge singularities. With certain assumptions regarding location of the nearest singularity to origin, $\mu_B=0$, one could estimate the location of the singularity from the Taylor coefficients. Furthermore, in the vicinity of a critical point, the LY singularities obey a particular scaling form. The key idea is to map the extrapolated singularity from the Taylor coefficients at different temperatures, referred to as the ``Lee-Yang trajectory", and compare it with the scaling form predicted by critical scaling. This approach allows the information extracted from the Taylor coefficients to extend beyond the region constrained by the truncated expansion. The idea of extracting the location of LY singularities from the truncated Taylor expansion goes back to work of Fisher \cite{Fisher:1974series}. The significance of the LY trajectory in the context of the QCD phase diagram and the critical point has been pointed out in \cite{Halasz_1997,Ejiri:2005ts,Stephanov:2006dn}. Recently, based on these observations, quantitative estimations of the LY trajectory were made  by extracting the radius of convergence or using Pad\'e type resummations \cite{Mukherjee:2019eou,Connelly:2020pno,Basar:2021hdf,Basar:2021gyi,Dimopoulos:2021vrk,Nicotra:2021ijp,Singh:2021pog,Schmidt:2022ogw,Bollweg:2022rps,Clarke:2023noy,Bollweg:2022fqq}.

In this paper, I employ an efficient method that improves Pad\'e resummation in order to construct the LY trajectory. I utilize the recent HotQCD results \cite{Bollweg:2022rps,Bollweg:2022fqq}  for the Taylor coefficients. From the improved resummation, assuming that the trajectory, in part, can be captured by the scaling behavior of the critical point, I extrapolate the location of the critical point and constrain the critical contribution to the equation of state in its vicinity. This is the main result of this paper and is summarized in Table \ref{table} in Section \ref{sec:results}. My hope is that these results can be incorporated into the state-of-the-art parametrizations of the equation of state such as in Refs. \cite{Parotto:2018pwx,An:2021wof} in to order to assist the experimental effort for the search for the critical point. In order to take into account lattice systematics, I also utilize data from the Wuppertal-Budapest \cite{Borsanyi:2018grb}. 

The rest of the paper is organized as follows. In Sec. \ref{sec:ly} I summarize the physics of LY singularities in the context of the QCD critical point and establish the notation that I use for the rest of the paper. In Sec. \ref{sec:resummations}, I build the necessary mathematical machinery that I use to extract the singularities from the Taylor coefficients. My results are presented in Sec. \ref{sec:results}. The same analysis is then repeated by using the Wuppertal-Budapest data in Sec. \ref{sec:systematics} and compared with the  main analysis that uses the HotQCD results. I extensively discuss these results in the final section, Sec. \ref{sec:discussion}.   

\section{Lee-Yang edge singularities }
\label{sec:ly}

In their seminal work in 1952, Lee and Yang showed that phase transitions of a thermodynamic system can be understood in terms of the complex singularities of the grand canonical partition function \cite{Yang:1952be,Lee:1952ig}. In general, the partition function, $Z(\z)$, of a system with finitely many degrees of freedom is a polynomial in fugacity, $\z=e^{\mu/T}$, and is nonnegative for $\z>0$.  Naturally, being a polynomial, it has zeros for complex values of $\z$, which, in the thermodynamic limit, coalesce into branch points that emanate from the so called Lee-Yang edge singularities. At a second order phase transition the LY edge singularities pinch the real axis.  

I will focus on the LY singularities associated with the three-dimensional Ising model which belongs to the same universality class as QCD.  In the Ising model, there are two relevant operators, energy density and spin, whose couplings are the reduced temperature, $r$, and the magnetic field, $h$. The critical point in the phase diagram sits at $r=h=0$. The transition between positive and negative magnetization phases (probed by $h>0$ and $h<0$ respectively) is a smooth crossover for $r>0$ and a first order transition for $r<0$. It is convenient to use the scaling variable $x=h r^{-\bd}$ to express the location the LY singularity which is simply along the pure imaginary axis:
\bea
x=\pm i x_{LY}\,.
\ea
Here $\beta$ and $\delta$ are the usual Ising critical exponents \cite{ZinnJustin:2002ru}. In this work I use the value computed from conformal bootstrap \cite{PhysRevD.86.025022}
\bea
\bd\approx 1.5631\,.
\label{eq:bd}
\ea 
The value of $x_{LY}$ was recently computed via functional renormalization group \cite{Rennecke:2022ohx,Connelly:2020gwa,Johnson:2022cqv} as well as using the Schofield representation \cite{Karsch:2023rfb}. Based on these works I will use the value:  
\bea
x_{LY}=|z_c|^{-\bd}\approx 0.246
\label{eq:xLY}
\ea
The LY singularity is a critical point as well and the equation of state in its vicinity belongs to the same universality class as the $\phi^3$ theory with a pure imaginary coupling \cite{Fisher:1978pf}. The universal, singular contribution to the magnetization around the LY singularity is given by 
\bea
m-m_c\sim (x\pm i x_{LY})^{\sigma_{LY}}
\ea  
with the critical exponent $\sigma_{LY}\approx0.074-0.085$ \cite{An:2016lni}.  
For a more detailed analysis of the analytical structure of the LY singularities in three and  two-dimensional Ising models I refer the reader to Refs. \cite{An:2016lni,An:2017brc} and \cite{Fonseca:2001dc} respectively. In the latter case, the physics of the LY singularity is captured by a non-unitary conformal field theory. In the former case, an analogous analytical construction does not exist at the moment, and one has to rely on numerical results such as the epsilon expansion.

The above observations about the three-dimensional Ising model can now be translated to QCD, given that these two theories are in the same universality class. In the vicinity of the critical point, $(T_c,\mu_c)$, the relevant directions in the QCD phase diagram, $T$ and $\mu_B$, can be mapped to those of the Ising model, $h$ and $r$, via a linear map \cite{Parotto:2018pwx,Pradeep:2019ccv} 
\bea
\label{eq:mapping}
\begin{pmatrix}
r\\ h
\end{pmatrix}:=
\begin{pmatrix}
r_T &&r_\mu 
\\
h_T   && h_\mu 
\end{pmatrix}
\begin{pmatrix}
T-T_c \\ \mu-\mu_c
\end{pmatrix}.\quad
\ea
This mapping then leads to the following expression for the trajectory of the LY singularities of QCD in the vicinity of the critical point \cite{Stephanov:2006dn}: 
\bea
\label{eq:ly-traj}
\mu_{LY}(T)\approx\mu_c -c_1(T-T_c) \pm i x_{LY} c_2 (T-T_c)^\bd, 
\nn
\text{where }c_1:={\hT\over \hmu}:=\tan\alpha_1 \quad c_2:={\rmu^\bd \over \hmu } \prt{{\rT\over\rmu}-{\hT\over\hmu} }^{\bd}.\quad
\ea
Notice that $c_1$ is the slope of the crossover line, 
whereas $c_2$ depends on the relative angle, $\alpha_2$,  between the $h$ and $r$ axes \cite{Parotto:2018pwx, Pradeep:2019ccv}. Remarkably, the trajectory in Eq. \eqref{eq:ly-traj} depends not only on the location of the critical point, but also on the non-universal mapping parameters. My goal is to reconstruct Eq. \eqref{eq:ly-traj}  and constrain these non-universal values from the Taylor series coefficients of the equation of state.

\section{Resummations and conformal maps}
\label{sec:resummations}

The main objective is to extract the singular behavior of the equation of state $p(T,\mu_B)$ from its Taylor series expansion around $\mu_B=0$,
\bea
\frac{p(T,\mu_B)-p(T,0)}{T^4}\approx \sum_{n=0}^N \frac{\chi_{2n}(T)}{(2n)!} \left( \mu_B\over T \right)^{2n}\,.
\label{eq:p-taylor}
\ea
In practice there is only access to a finitely many terms, and the truncated Taylor series is a polynomial which obviously has no singularities. However even with finitely many coefficients, one could extract the singular behavior of the equation of state. Assuming that the nearest singularity to the origin is the LY singularity, following the extended analyticity conjecture of Fonseca and Zamalodchikov  \cite{Fonseca:2001dc}, along with Darboux's theorem, it is possible to estimate both $|\mu_{LY}|$ via the radius of convergence by using the standard root/ratio tests \cite{Bazavov:2017dus,GIORDANO2021121986,Mukherjee:2019eou}, as well as $\arg \mu_{LY}$ \cite{Basar:2021hdf,Basar:2021gyi}. However, here, the leading singularities are in fact a complex conjugate pair [see Eq. \eqref{eq:ly-traj}] and consequently the roots/ratios of the coefficients exhibit oscillatory behavior due to interference between the phases of the two singularities which render these estimations numerically challenging. 
 
A better approach is to use Pad\'e resummation, which approximates the original function, $p(\mu_B^2)$\footnote{To keep the notation compact, I suppress the temperature argument, and switch the argument of $p$ to $\mu_B^2$.}, by a rational function, 
 \bea
 {\rm P}[p](\mu_B^2):=\frac{q_1(\mu_B^2)}{q_2(\mu_B^2)}
 \label{eq:pade}
 \ea
where $p_1$ and $p_2$ are polynomials of order $[N/2]$\footnote{
 In this work I only consider diagonal Pad\'e approximants where the polynomials $p_1$ and $p_2$ are of the same order.}, whose coefficients are determined by Taylor expanding $ {\rm P_{N/2}}[p](\mu_B^2)$ and matching the coefficients with the original ones in Eq. \eqref{eq:p-taylor}.\footnote{Note that even though the coefficients of these polynomials depend on $T$, they are not necessarily smooth functions of $T$.} The underlying singularities of $p(\mu_B^2)$ are then represented by accumulation points the poles and zeros of $ {\rm P}[p](\mu_B^2)$. Furthermore, if the underlying singularity is a branch point, the poles/zeros lie on curved arcs.  Remarkably, the shapes of these arcs can be understood in terms of a  two-dimensional electrostatic problem of finding a configuration of charges (poles and zeros) which minimizes an effective capacitance \cite{STAHL1997139, saff, Costin:2020pcj}. At the same time, when there is a complex conjugate pair of singularities, as in QCD, the arcs that emerge from each singularity coalesce along the real axis. These singularities along the real axis are unphysical, but they are not numerical artifacts; their existence is unavoidable with Pad\'e resummation.  To make things worse, their number grows as $N$ increases, limiting the applicability of the Pad\'e resummation to $\mu\lesssim |\mu_{LY}|$.

In order to overcome this inherent shortcoming of Pad\'e resummation, the next step is to pair it with a conformal map. In a way, Pad\'e resummation already introduces its own conformal map by representing the branches with the curved arcs that minimize the effective capacitance. It is possible to further improve convergence properties of Pad\'e, and extend its domain of applicability by using an additional conformal map. The idea is to map the complex $\mu_B^2$  plane, where the equation of state is expressed in, to a different (preferably compact) region, such as the unit disk, denoted by $\z$ via a conformal map,
\bea
\mu_B^2:=\phi(\z)\,.
\ea
 The next step is to expand the equation of state as a series expansion in $\z$ and perform the Pad\'e resummation for this expansion:
 \bea
 {\rm CP}[p](\z):=\frac{\tilde q_1(\z)}{\tilde q_2(\z)}\,.
 \ea
 where $\tilde p$ and $\tilde q$ are order $N/2$ polynomials whose coefficients are determined by the Taylor coefficients of $p(\phi(\z))$. This resummation will be referred to simply as ``conformal Pad\'e''. Different choices of the conformal map, $\phi$, leads to significant improvements over the usual Pad\'e resummation. Moreover, extra information such as the singular behavior of the original function can be ``baked in" into the resummation via choosing an appropriate conformal map. Note that, in contrast, the only input of Pad\'e resummation is the set of Taylor coefficients.
 After performing the Pad\'e resummation in the $\z$ plane, the original function is then represented as
 \bea
f(\mu_B^2) \underset{\text{conf. Pad\'e}}{=} {\rm CP}[p](\z)\big|_{\zeta=\phi^{-1}(\mu_B^2)}\,.
 \ea
 For certain conformal maps, an analytical expression for the inverse function exists, and for others it does not and the inversion has to be computed numerically. Similar to Pad\'e, the original singularities of the function are represented by accumulation of zeros and poles of conformal Pad\'e in $\z$ plane. They can be mapped back to $\mu_B^2$ plane via $\phi(\zeta)$. Note that the computation of the singularities does not require the inverse function. The key point is that for the conformal maps that are considered here, the spurious poles appear \textit{outside} the unit disk and therefore are not present in the $\mu_B^2$ plane \cite{Costin:2020hwg,Costin:2021bay}. The conformal maps that are used in this work are discussed below.

\begin{figure}[h]
\includegraphics[scale=0.55]{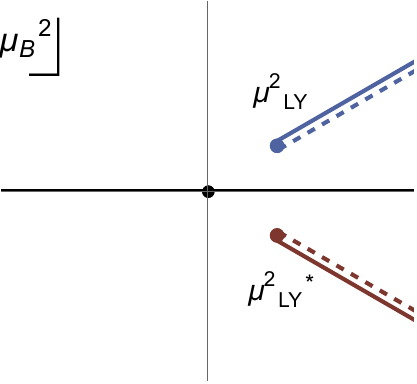}\qquad
\includegraphics[scale=0.55]{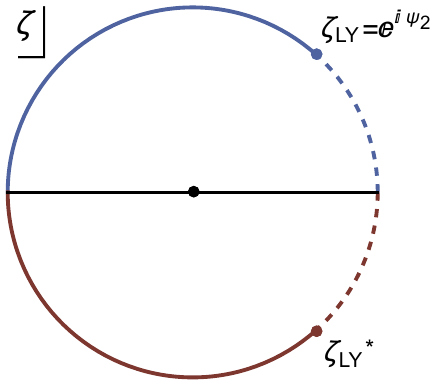}
\caption{The original $\mu_B^2$ plane (right) with radial cuts and its image in the unit circle (right) under the two-cut map defined in Eq. \eqref{eq:phi2}.}
\label{fig:2cut}
\end{figure}

The first map, named as the``two-cut map", is defined as
\bea
\label{eq:phi2}
\phi_2(\z)=4|\mu^2_{LY}|\zeta\left[\frac {\theta/\pi}{(1-\z)^2}\right]^{\theta/\pi}\left[\frac {1-\theta/\pi}{(1+\z)^2}\right]^{1-\theta/\pi}\,.
\ea
It maps the complex plane $\mu_B^2$ plane with two radial cuts into the unit disk as shown in Fig. \ref{fig:2cut}. The branch points located at $\mu_B^2:=|\mu_{LY}^2|e^{\pm i \theta}$ are mapped to the edge of the unit disk with the angle $\psi_2:=2\arcsin(\sqrt{\theta/\pi})$:
\bea
|\mu_{LY}^2|e^{\pm i\theta} \rightarrow \zeta_{LY}=e^{\pm i\psi_2} 
\ea
and the branch cuts are mapped to the edge of the unit disk as show in Fig. \ref{fig:2cut}.  Notably, this map has been used in other physical applications, mostly within the context of extracting the Borel singularities associated with asymptotic series  \cite{Guida:1998bx,Rossi:2018,Serone:2019szm,Costin:2020hwg,Costin:2021bay}. As opposed to Pad\'e, conformal Pad\'e with the two-cut map does not generate unphysical poles along the real axis, allowing one to reconstruct the original function \textit{beyond} the radius of convergence. Furthermore it generally gives a better estimate for the location of the branch singularities compared to Pad\'e with the same number of Taylor coefficients \cite{Costin:2020pcj}. 
One shortcoming of the two-cut map, however, is that its applicability is limited to the first Riemann sheet. Namely, even though it gives a good estimate for the location of the branch singularities, it does not allow one to actually go through the branch cuts. This can be seen from the fact that it maps the first Riemann sheet, bounded with the radial branch cuts, inside the whole unit disk.  

An improvement over the two-cut map is the ``uniformizing map" defined as
\bea
\label{eq:phiu}
\phi_u(\tau)=
|\mu^2_{LY}|
\left(e^{-i\theta} +2 i \sin\theta\lambda\left( \tau \right)\right)\,.
\ea
Here  $\lambda(\tau)=\theta_2^4(\tau)/\theta_3^4(\tau)$ is the modular lambda function with $\theta_2(\tau)=\sum_{n=\infty}^\infty e^{2\pi i \tau (n+1/2)^2}$ and $\theta_3(\tau)=\sum_{n=\infty}^\infty e^{2\pi i \tau n^2}$ being the usual Jacobi elliptic functions defined in the upper half plane $\im \tau>0$. I further map the upper half-plane into the unit disk via the following Mobius transformation:
\bea
\label{eq:mobius}
\tau\rightarrow\tau(\z)=i\frac{\K\left(\frac12-\frac i2\cot\theta\right) +  \K\left(\frac12+\frac i2\cot\theta\right) i\z}{\K\left(\frac12+\frac i2\cot\theta\right) -  \K\left(\frac12-\frac i2\cot\theta\right)i \z}
\ea
 where $\K(m)$ is the complete elliptic integral of the first kind:
\bea
\K(m)=\int_0^{\pi/2} \frac{d\theta}{\sqrt{1-m\sin^2\theta}}\,.
\label{eq:k}
\ea
This map ``uniformizes" the multi-sheeted $\mu_B^2$ plane of the original function by mapping the \textit{entire} multi-sheeted domain into a simply connected domain. A path that crosses the branch cuts in the $\mu_B^2$  plane is represented by a smooth curve in the mapped plane. How this works can be briefly explained in two steps. 

\begin{figure}[h]
\includegraphics[scale=0.55]{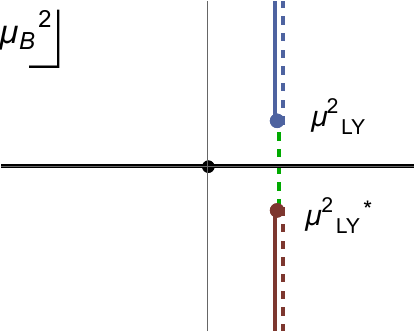}\quad
\includegraphics[scale=0.53]{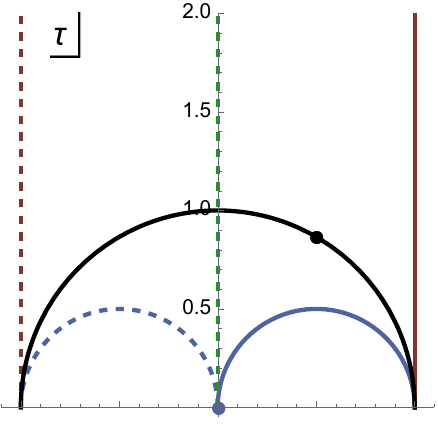}\quad
\includegraphics[scale=0.55]{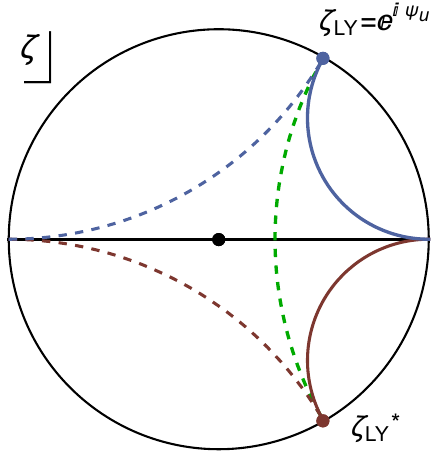}
\caption{The original $\mu_B^2$ plane (right) with vertical cuts and its image in the modular $\tau$ plane (center) and the unit circle (right) under the uniformizing map defined in Eqs. \eqref{eq:phiu} and \eqref{eq:mobius}.}
\label{fig:uni}
\end{figure}

The first step is to map the $\mu_B^2$ plane with vertical branches into the fundamental domain via Eq. \eqref{eq:phiu}. This map is shown in Fig. \ref{fig:uni} (center). Notice that the entire $\mu_B^2 $ is mapped into a subset of the upper half-plane, the fundamental domain. The remaining half-circular ``gaps" are filled by Schwartz reflection of the fundamental domain. These reflections are transformations built out of the elementary modular transformations, $S:\tau-\rightarrow -1/\tau$ and $T:\tau\rightarrow\tau+1$. Each Schwartz reflection fills in a portion of the semi-circular gaps in the upper half-plane, a process that can in principle continued ad-infinitum, asymptotically filling the whole gap. These regions are the images of the higher Riemann sheets with each Schwartz reflection corresponding to a particular sheet. Therefore, moving across different sheets in the $\mu_B^2$ plane is simply represented by moving smoothly in the $\tau$ plane. Each Schwarz reflection generates a progressively smaller region in the $\tau$ plane. Therefore, in order to resolve the higher sheets one needs progressively more precise information, meaning more Taylor coefficients.  

The second step is to further map the fundamental domain into the unit disk via the Mobius transformation given in Eq. \eqref{eq:mobius}. As seen in Fig. \ref{fig:uni} (right), the branch points are mapped to the edge of the unit disk. Similar to the two-cut case, the angle is given by
\bea
|\mu_{LY}^2|e^{\pm i \theta}\rightarrow \z_{LY}= i\frac{\K\left(\frac12\mp \frac i2 \cot\theta\right)}{\K\left(\frac12\pm \frac i2 \cot\theta\right)} :=e^{\pm i\psi_u}
\ea
The branches, however are mapped into certain curves that for a boundary of a compact domain within the unit circle. This compact, skewed diamond shaped region bounded by these curves is image of the first sheet. The remaining gaps in the unit circle are filled with Mobius transformations of the Schwartz reflected regions in the modular $\tau$ plane. These are the images of higher sheets. Consequently the entire multi sheeted $\mu_B^2$ space is mapped into the unit circle and  crossing between is represented by smooth paths within the unit circle. 

This way, one can reconstruct the original function not only beyond its radius of convergence, but also in the higher Riemann sheets, using only the Taylor coefficients of the expansion around the origin in the first Riemann sheet. In the context of this work, the higher Riemann sheets encode the scaling equation of state in the first order phase transition region, $T<T_c$. In Ref. \cite{Basar:2021gyi} I demonstrated this by reconstructing the equation of state of the mean field Ising model in the first-order transition region ($T<T_c$) by using the Taylor expansion coefficients of the high-temperature ($T>T_c$) equation of state obtained in the first Riemann sheet. In addition to reconstructing the equation of state in the higher sheets, the uniformizing map also provides a better approximation to the function compared to Pad\'e and two-cut conformal Pad\'e. 

This work will focus on the first sheet and  the uniformizing map will be uses to get a better estimate for the LY singularities and the equation of state. This is because having access to only four Taylor coefficients with sizable statistical uncertainties makes the analytical continuation to higher sheets extremely challenging. As opposed to the two-cut map, there is an analytical expression for the inverse of the uniformizing map, given as,
\bea
\phi_{uni}^{-1}(\mu_B^2 )=-i\frac{\K\left(\kappa_+\right) \K\left(\kappa_-+ \frac{i}{2\sin\theta}\frac{ \mu^2_B}{|\mu^2_{LY}|} \right)-\K\left(\kappa_-\right) \K\left(\kappa_+ - \frac{i}{2\sin\theta}\frac{ \mu^2_B}{|\mu^2_{LY}|} \right)}
{\K\left(\kappa_-\right) \K\left(\kappa_-+ \frac{i}{2\sin\theta}\frac{ \mu^2_B}{|\mu^2_{LY}|} \right)+\K\left(\kappa_+\right) \K\left(\kappa_+ - \frac{i}{2\sin\theta}\frac{ \mu^2_B}{|\mu^2_{LY}|} \right)}
\ea
where $\kappa_\pm:=(1\pm i\cot\theta)/2$ and $\theta=|\arg \mu^2_{LY}|$. 

Notice that both the two-cut map and the uniformizing map depend on the location of the singularity, $\mu_{LY}$ which, in fact, is what one wants to extract from these resummation techniques. This seemingly paradoxical situation can be overcome by a simple procedure: 1) guess the location from ordinary Pad\'e and use this initial guess as the value of $\mu_{LY}$ in the conformal map. 2) Extract the poles from conformal Pad\'e in $\z$ plane whose images in the $\mu_B^2$ plane gives a refined estimate for $\mu_{LY}$. 3) Iterate the same procedure by using this refined estimate for the value of $\mu_{LY}$ in the conformal map. I observed that this iteration converges to a value which constitutes the final estimate for $\mu_{LY}$. This iterative process is used for different temperatures to construct the LY trajectory. 

Before presenting the results it is worth to comment on an alternative resummation technique. The equation of state in the vicinity of the LY singularities has a singular contribution, but this singular contribution actually vanishes at the singularity for the pressure and generates a cusp for the density. However, a true divergence occurs for the susceptibility. For real values of $\mu_B$ and $T\gtrsim T_c$, the susceptibility does not diverge but sharply peaks around $\re\mu_{LY}$. For this reason, sometimes performing a Pad\'e resummation for the susceptibility instead of the pressure can give a better estimate of the underlying singularities as well as the function itself. Therefore I also performed resummation for the susceptibility:
\bea
\chi(T,\mu_B)={\del^2 p\over \del \mu_B^2}\approx \sum_{n=0}^{N-1}  \frac{ \chi_{2n+2}(T)}{(2n)!}\left(\mu_B\over T\right)^{2n}\,.
\label{eq:chi-taylor}
\ea
Using pressure or susceptibility in the resummation just amounts to reshuffling the same Taylor coefficients. However this reshuffling, especially when $N$ is small, does make a difference. For example in Refs. \cite{Basar:2021hdf, Basar:2021gyi} which studied the Gross-Neveu and the Chiral Random Matrix models, using the susceptibility led to much more accurate results than the pressure. For QCD, this situation is more complicated due to the statistical uncertainties in the coefficients as will be discussed further in the next section. 

\section{Results}
\label{sec:results}

 The results for estimates for the Lee-Yang singularities as a function of temperature, obtained by the resummation methods are presented in this section. The calculations for the locations of the critical point and the non-universal mapping parameters in Eq. \eqref{eq:mapping}
 based on these estimates are presented as well, along with results for the susceptibility as a function of chemical potential for a few different temperatures.

\begin{figure}[h]
\includegraphics[scale=0.5]{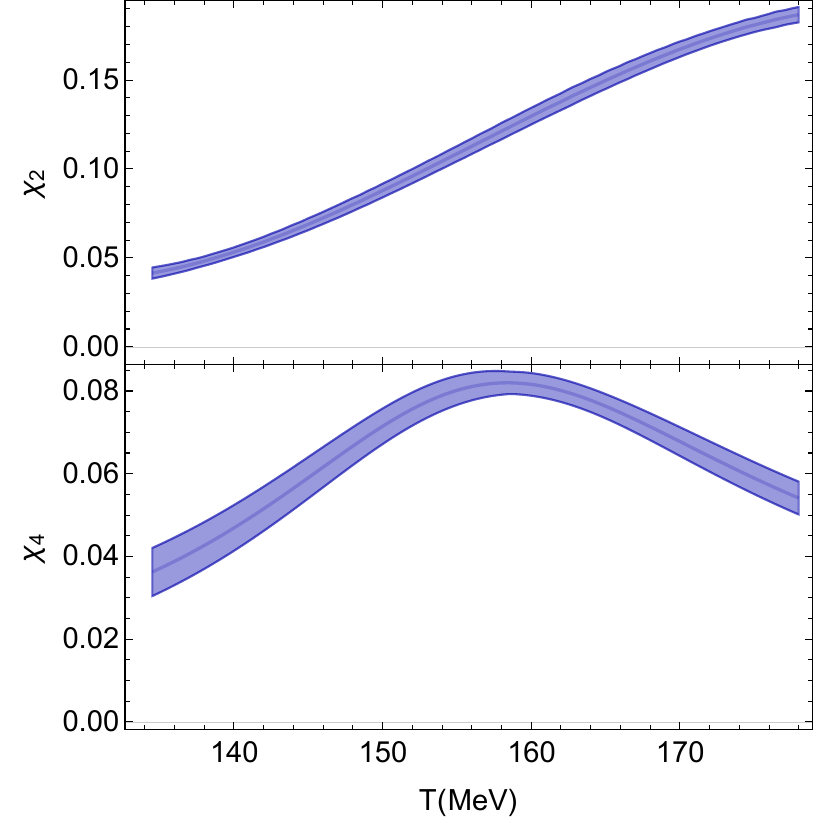}
\includegraphics[scale=0.5]{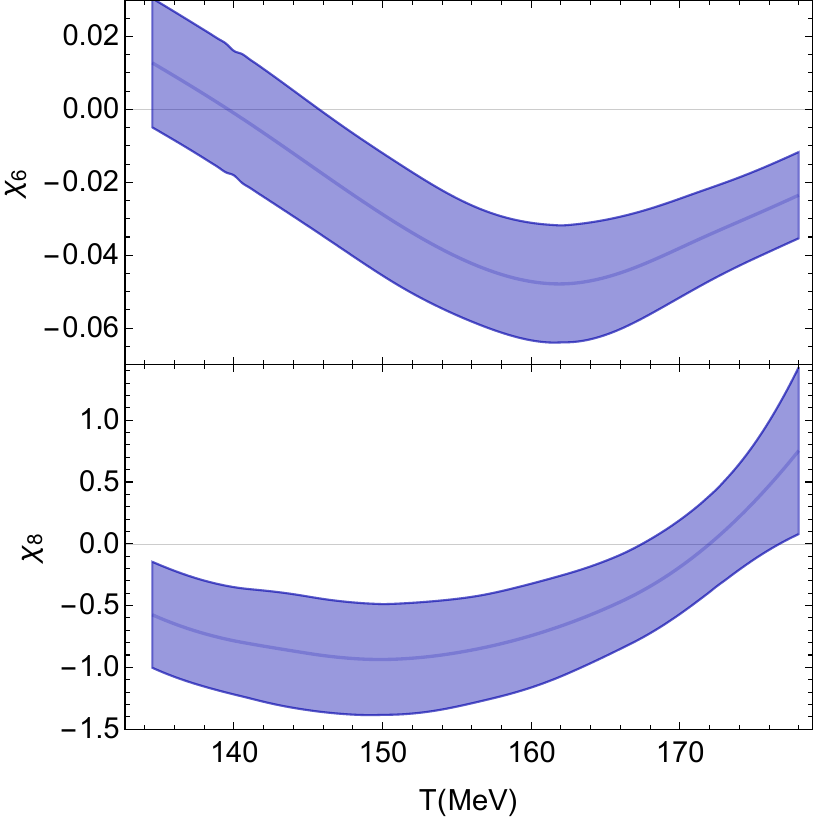}
\caption{The  HotQCD data for Taylor coefficients of the equation of state, Eq. \eqref{eq:p-taylor}, for $\mu_Q=\mu_S=0$ from Ref. \cite{Bollweg:2022rps}. The bands denote the continuum extrapolation for $\chi_2$ and $\chi_4$ and spline extrapolation from the $N_\tau=8$ data for $\chi_6$ and $\chi_8$.}
\label{fig:chis}
\end{figure}

I use the recent results for the Taylor coefficients up to ${\cal O}(\mu_B^8)$ computed by the HotQCD collaboration \cite{Bollweg:2022rps} for $\mu_S=\mu_Q=0$. The continuum extrapolation for the first two terms, $\chi_2$, $\chi_4$ are used. Unfortunately, continuum extrapolations for the remaining two terms, $\chi_6$ and $\chi_8$, were not available at the time this work was completed. I use the spline extrapolation for the $N_\tau=8$ data given in Ref. \cite{Bollweg:2022rps}.  For sake of completeness the HotQCD data are plotted in Fig. \ref{fig:chis}.

With four terms in the Taylor expansion, the diagonal Pad\'e resummation is a ratio of two quadratic polynomials in $\mu_B^2$.  Analytical expressions for the poles and zeros of Pad\'e \cite{Bollweg:2022rps} as well as conformal Pad\'e can be obtained; however, their functional forms do not play a central role in this analysis and to keep the discussion concise they will not be included here.  

\begin{figure}[h]
\includegraphics[scale=0.5]{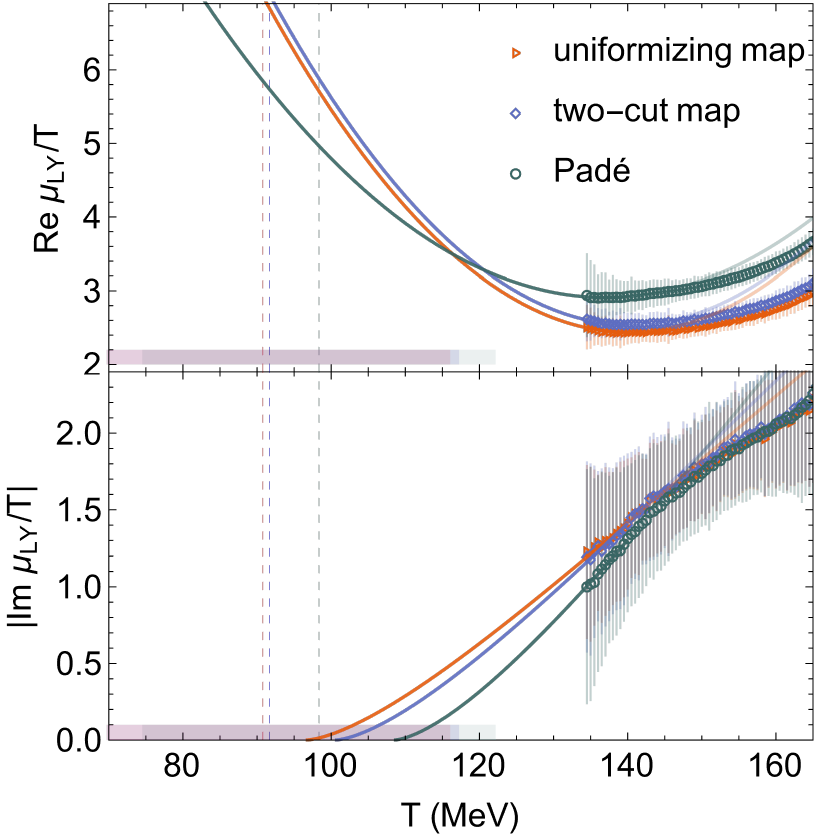}\quad
\includegraphics[scale=0.5]{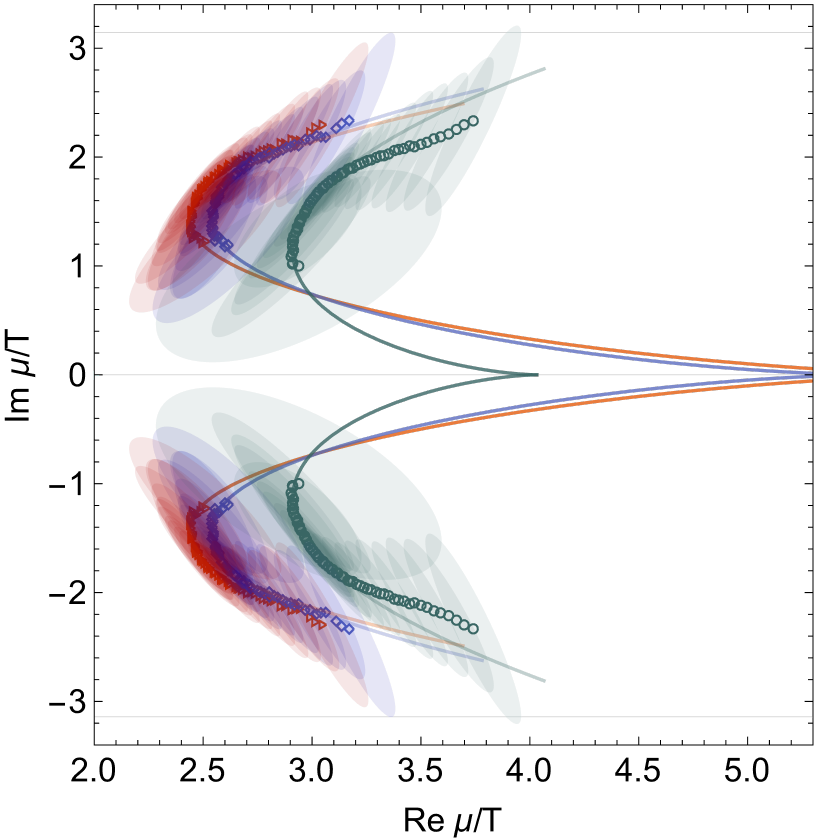}
\caption{ The Lee-Yang trajectory constructed via different resummations. The solid lines denote the best fits for real and imaginary parts given in Eq. \eqref{eq:fits}. The dashed vertical lines denote the best fit extrapolation to $\mu_c^2$. Where the fit for the real part intersects the dashed lines leads to the extrapolation of $T_c$. The bars and ellipses on the left/right figures represent $1\sigma$ uncertainty stemming from the noise in the Taylor coefficients.}
\label{fig:LY-traj}
\end{figure}

In order to take into account the statistical uncertainties, I sampled an ensemble of coefficients from a Gaussian distribution and used diagonal (2,2) Pad\'e resummation which has a complex conjugate pair of poles.  Since there are no other poles to form an accumulation point, these poles are used as estimators for the LY singularities, $|\mu_{LY}^2|e^{\pm i\theta}$. 
The LY trajectory constructed by repeating these steps for different temperatures is shown in Fig. \ref{fig:LY-traj}. I also found that the zeros did not follow any meaningful pattern. This is likely due to $N$ being relatively small. 

\begin{figure}[h]
\includegraphics[scale=0.45]{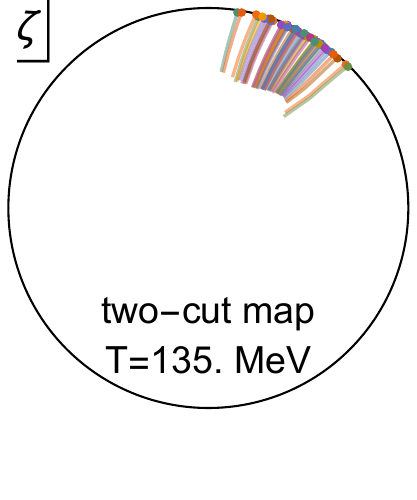}\quad
\includegraphics[scale=0.49]{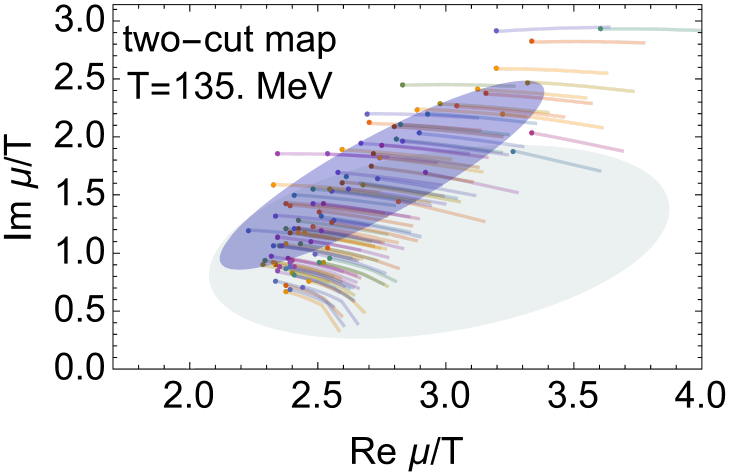}\quad
\includegraphics[scale=0.49]{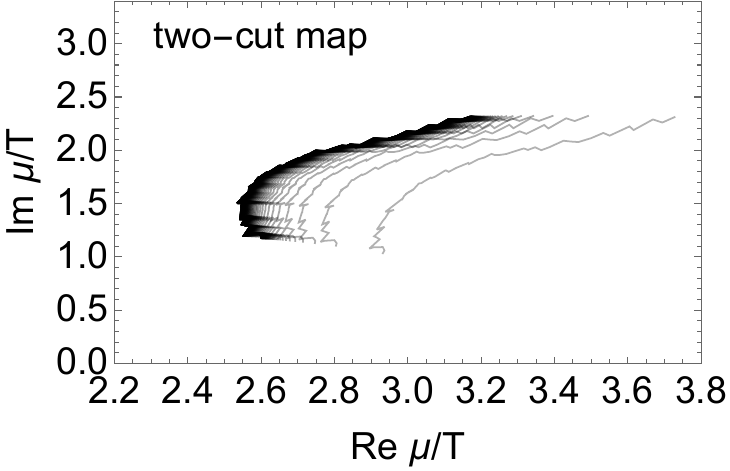}
\caption{ The iterative procedure of extracting the singularities via the two-cut conformal map. Each line represent the trajectory that corresponds to a set of Taylor coefficients sampled form a Gaussian ensemble.  The solid blue (green) disks denote the $1\sigma$ uncertainty region for two-cut and ordinary Pad\'e resummations.  Further description is in the text.}
\label{fig:iteration-2cut}
\end{figure}

\begin{figure}[h]
\includegraphics[scale=0.45]{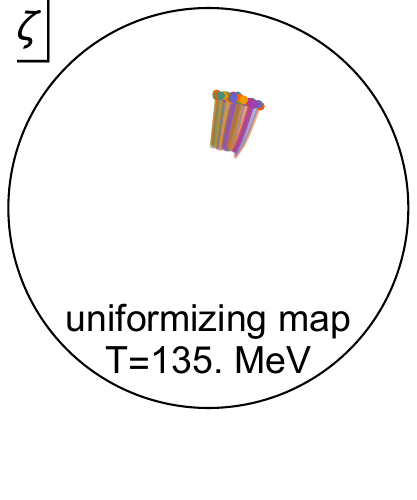}\quad
\includegraphics[scale=0.49]{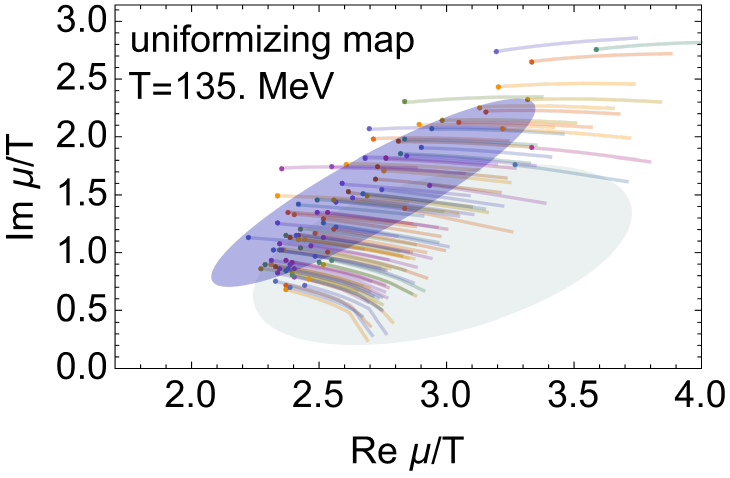}\quad
\includegraphics[scale=0.49]{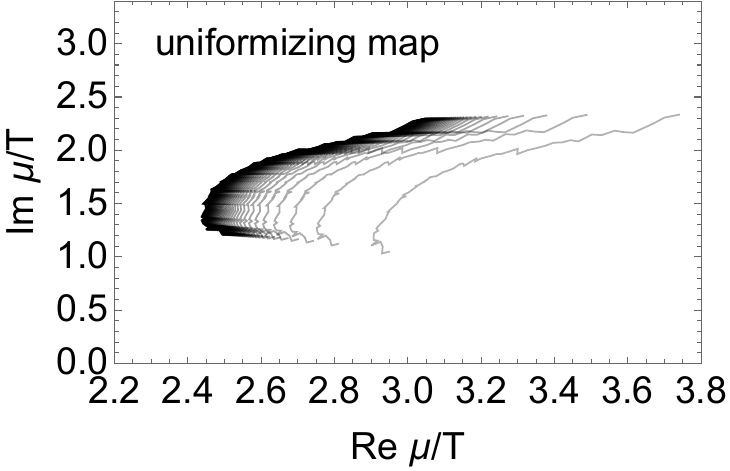}
\caption{ The iterative procedure of extracting the singularities via the uniformizing conformal map. Each line represent the trajectory that corresponds to a set of Taylor coefficients sampled form a Gaussian ensemble. The solid blue (green) disks denote the $1\sigma$ uncertainty region for uniformizing and ordinary Pad\'e resummations. Further description in text.}
\label{fig:iteration-uni}
\end{figure}

With the conformal maps, there is one more step in extracting the singularity. As mentioned in the previous section, the conformal maps explicitly depend on the location of the singularity. I first used the Pad\'e estimate for the singularity in the conformal map and then followed the iterative procedure described in Section \ref{sec:resummations} to refine the estimate for $|\mu_{LY}^2|e^{\pm i\theta}$. The results of this procedure are shown in Figs. \ref{fig:iteration-2cut} and Fig. \ref{fig:iteration-uni}.

\begin{figure}[h]
\includegraphics[scale=0.43]{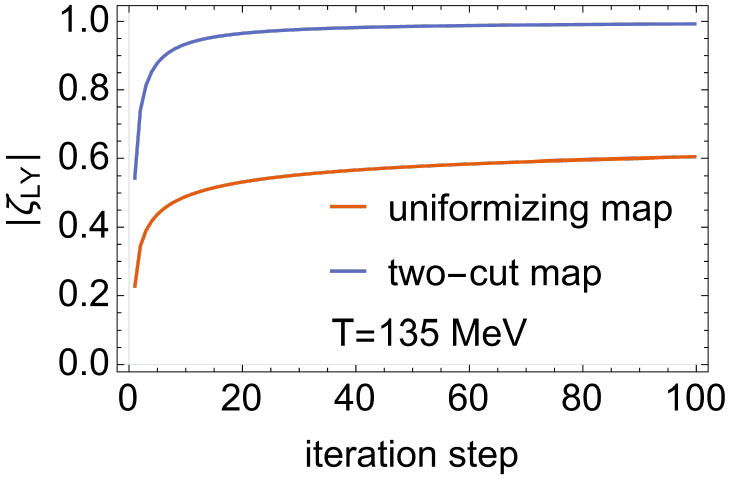}\,\,
\includegraphics[scale=0.43]{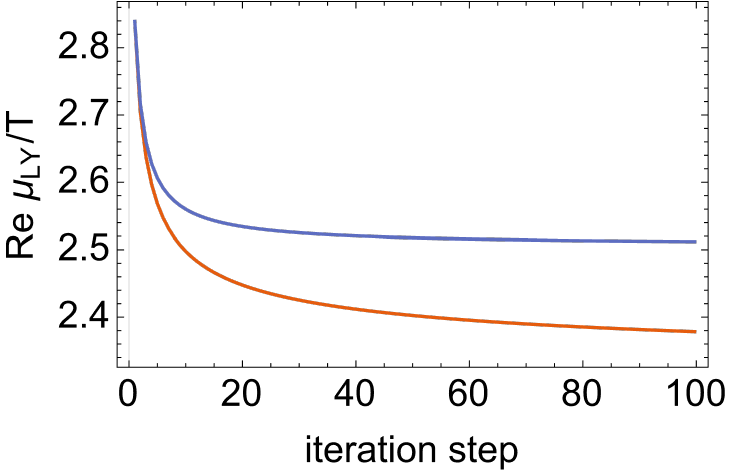}\,\,
\includegraphics[scale=0.43]{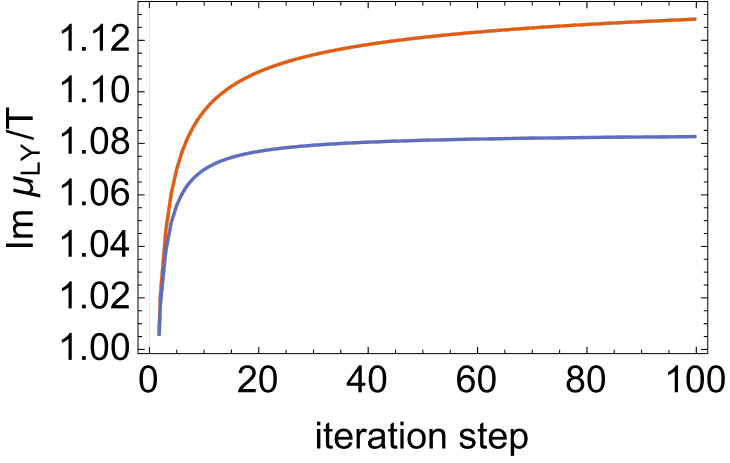}
\caption{The convergence of $\mu_{LY}$ with the iteration procedure. The left figure shows the convergence of the absolute value of the image of $\mu_{LY}$ towards the edge of the unit disk and the center/right figures show $\mu_{LY}$. These plots represent a single line in Figs. \ref{fig:iteration-2cut} and \ref{fig:iteration-uni} (left/center). }
\label{fig:iteration}
\end{figure}

Figure \ref{fig:iteration-2cut} shows the trajectory of the iteration for different set of Taylor coefficients sampled from a Gaussian ensemble. Each line with color represents the iteration obtained with fixed set of Taylor coefficients sampled from a Gaussian ensemble, in the $\z$ (left) and $\mu_B$ (right) planes. The final estimate for each trajectory is denoted by a small solid disk. Each disk and the iteration curve are color coded. The pale blue disk in the center figure represents $1\sigma$ uncertainty region. For comparison, the $1\sigma$ uncertainty region for the ordinary Pad\'e resummation with the same temperature and ensemble of Taylor coefficients is also included. The right figure shows the evolution of the LY trajectory constructed by averaging over the Gaussian ensemble with the iteration. Different opacities denote different steps in the iteration, darker being later. Recall that the image of the true singularity of the equation of state in the $\z$ plane is along the unit disk (see Figs. \ref{fig:2cut} and \ref{fig:uni}). Remarkably the iteration indeed converges to the edge of the unit disk. Fig. \ref{fig:iteration-uni} shows the same trajectories for the uniformizing map. The same ensemble of Taylor coefficients is used both for ordinary Pad\'e and the two conformal maps. In Fig. \ref{fig:iteration} the result of the iteration for a single trajectory is shown. For both conformal maps, the initial point is the same Pad\'e pole obtained from the Gaussian ensemble. The left figure show the convergence towards the edge of the unit disk. 

The LY trajectories constructed from different resummations are shown in Fig. \ref{fig:LY-traj}. Each point for conformal Pad\'e is obtained after 100 iterations as described above. The error bars represent the $1\sigma$ uncertainties inherited from the statistical uncertainties in the Taylor coefficients. 

The next goal is to extract the location of the critical point from the LY trajectory. The critical chemical potential, $\mu_c$ is the point where $\im \mu_{LY}$ vanishes which is clearly beyond the available data as $\im \mu_{LY}>0$ for the temperature range in hand. At the same time, the fact that $\im \mu_{LY}$ is decreasing with decreasing temperature is suggestive that if there is a critical point, it lies at $T_c<135$ MeV. By extrapolating the trajectory to the point where $\im \mu_{LY}=0$ I estimate $\mu_c$ and $T_c$ using the following fits for the extrapolation
\bea
\re \mu_{LY}(T)&=& a_0+a_1(T-T_c)+a_2(T-T_c)^2\\
\im \mu_{LY}(T) &=& a (T-T_c)^{\beta\delta}
\label{eq:fits}
\ea
whose form is motivated by the scaling form given in Eq. \eqref{eq:ly-traj}. The results for the best-fits for different resummations are shown in Fig. \ref{fig:LY-traj} as solid lines. In these fits, I used the first 20 terms in the trajectory with $134.5 \leq T \leq 144$. Finally, the location of the critical point, as well as the non-universal mapping parameters, the slope of the crossover line at the critical point, $\alpha_1$, and $c_2$ as given in. Eq. \eqref{eq:ly-traj}, extrapolated from these fits, are listed in Table \ref{table}. 

\begin{table}
\renewcommand{\arraystretch}{1.7}{\begin{tabular}{  | c  | c | c | c |  c | } 
 \hline
 \, & $T_c$ (MeV) & $\mu_c $ (MeV) & crossover slope ($\alpha_1$) & $c_2 $ (MeV$^{1-\beta\delta}$)
 \\
\hline
uniformizing  \,&\,  $97^{+18}_{-18}$  \,&\,  $ 579^{+172}_{-160}$  \,&\,    $9.40^\circ\,^{+3.89}_{-3.81}$ \,&\,  $2.22^{+0.52}_{-0.86}$ \,
\\  
 two-cut \,&\, $ 100^{+18}_{-18}$   \,&\,  $557^{+175}_{-150}$  \,&\,   $8.69^\circ\,^{+3.91}_{-3.83}$ \,&\, $2.56^{+0.58}_{-1.21}$ \,
%\\ 
%1-cut \,&\,   105 \,&\,  493   \,&\,   $6.63^\circ$  \,&\,  2.84 \,
\\  
Pad\'e \,&\,  $108^{+21}_{-21}$  \,&\,  $ 437^{+114}_{-50}$  \,&\,    $4.55^\circ\,^{+3.41}_{-3.37}$ \,&\,  $3.35^{+0.82}_{-1.37}$ \,
\\
 \hline
 \end{tabular}
 }
 \caption{The location of the critical point and the Ising model mapping parameters given in Eq. \eqref{eq:ly-traj} extracted from Pad\'e and conformal Pad\'e. The sub/superscripts denote the $1\sigma$ uncertainty}
 \label{table}
\end{table}

\begin{figure}[h]
\includegraphics[scale=0.33]{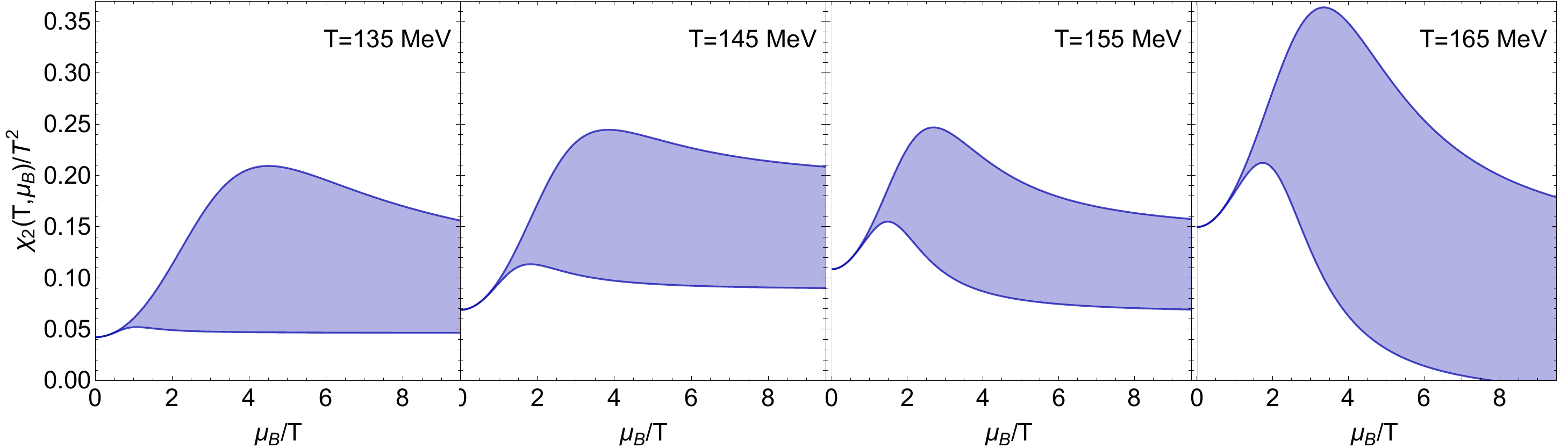}
\caption{The susceptibility as a function of $\mu_B$ for four different temperatures calculated via the uniformizing conformal Pad\'e from the Taylor coefficients of the susceptibility given in Eq. \eqref{eq:chi-taylor}.}
\label{fig:sus}
\end{figure}

\begin{figure}[h]
\includegraphics[scale=0.5]{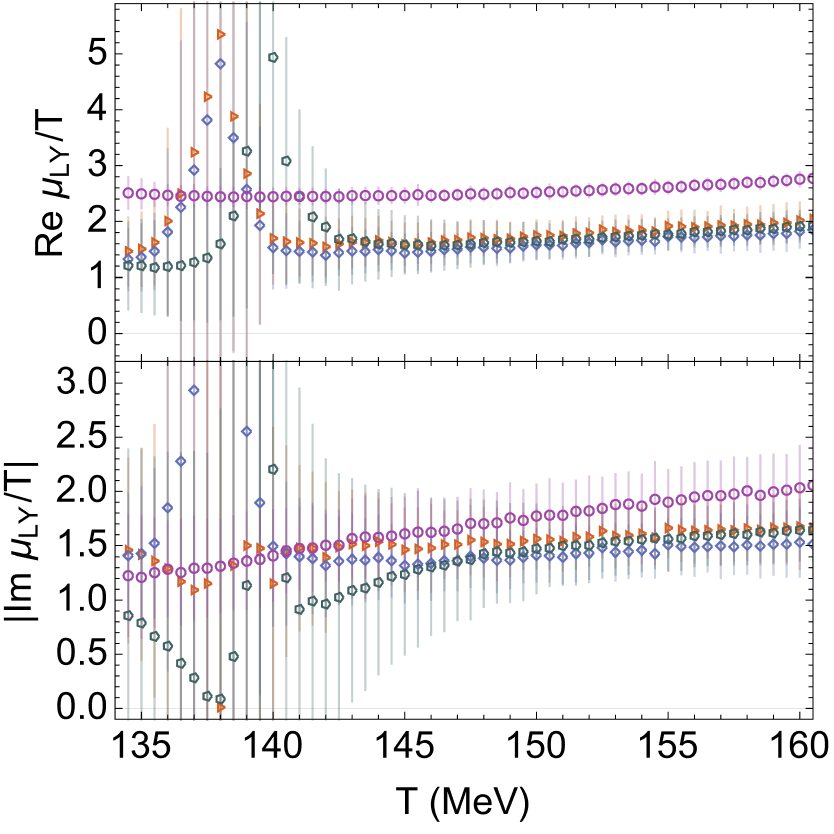}
\includegraphics[scale=0.45]{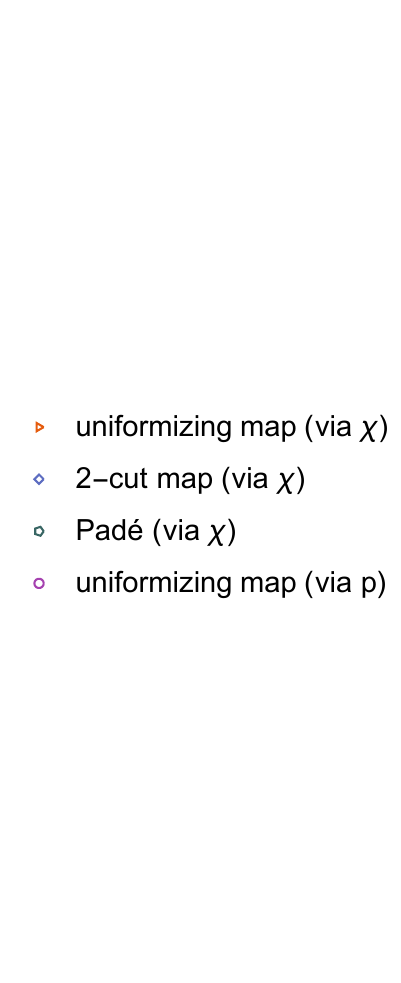}
\caption{The poles of Pad\'e and conformal Pad\'e extracted from the Taylor coefficients of susceptibility compared with those extracted from the pressure.}
\label{fig:LY-traj-compare}
\end{figure}

The final set of results concerns the Pad\'e and conformal Pad\'e resummations by using the Taylor coefficients of the susceptibility given in Eq. \eqref{eq:chi-taylor}. A particularly interesting result obtained from this resummation is the susceptibility as a function of chemical potential, shown in Fig. \ref{fig:sus}. The band denotes the $1\sigma$ uncertainty as before. Finally Fig. \ref{fig:LY-traj-compare} shows the singularities obtained this way. For comparison, uniformizing map results for the singularities obtained from the expansion of pressure are also shown in the same figure (in purple). 

Notice that using the coefficients of susceptibility reproduces qualitatively the expected behavior of $\chi$ as a function $\mu_B$, namely exhibiting a peak at some nonzero value of $\mu_B/T$, albeit with sizable statistical uncertainty. At the same time, the singularities extracted this way are less reliable compared to using the coefficients of pressure. I  found that the conformal Pad\'e singularities in the $\z$ plane were much further away from the edge of the unit disk than those obtained from the coefficient of pressure. Furthermore the statistical errors are very large, especially for $T\lesssim 145$ MeV. This is ultimately because the Pad\'e and conformal Pad\'e singularities depend rather sensitively on $\chi_6$, which changes sign around $140$ MeV, making the relative error quite large.  Due to the difference in the functional dependence of the Pad\'e singularities in the coefficients, this is not the case for the expansion of pressure. Therefore I conclude that using coefficients of pressure in the estimates for the location of singularities, using pressure as opposed to susceptibility is more reliable. However, using the susceptibility seems to reproduce the qualitatively expected form of the equation of state better. A more detailed analysis of this observation is left for future work. 

\section{Systematics}
\label{sec:systematics}

In addition to the statistical uncertainties that are present in any stochastic computation, and the uncertainties in extracting the singularities due to finite number of Taylor coefficients, there are also systematic uncertainties inherited from the lattice computations.The statistical uncertainties are already quantified in my analysis. The uncertainties due to the finite number of Taylor coefficients are more difficult to quantify but they can be gauged by analyzing how close the images of the extrapolated singularities are to the edge of the unit disk in the conformal plane, as explained in the previous section. At the same time, the systematic uncertainties are much more challenging to quantify. In order to shed light on these systematic uncertainties, in this section the same analysis described above is repeated by using a different data set of Taylor coefficients computed by the Wuppertal-Budapest (WB) Collaboration \cite{Borsanyi:2018grb}. The WB simulations are done at imaginary chemical potential with the same physical volume as HotQCD, in particular on $48^3\times 12$ (WB) and  $32^3\times8$ (HotQCD) lattices. 

\begin{figure}[h]
\includegraphics[scale=0.48]{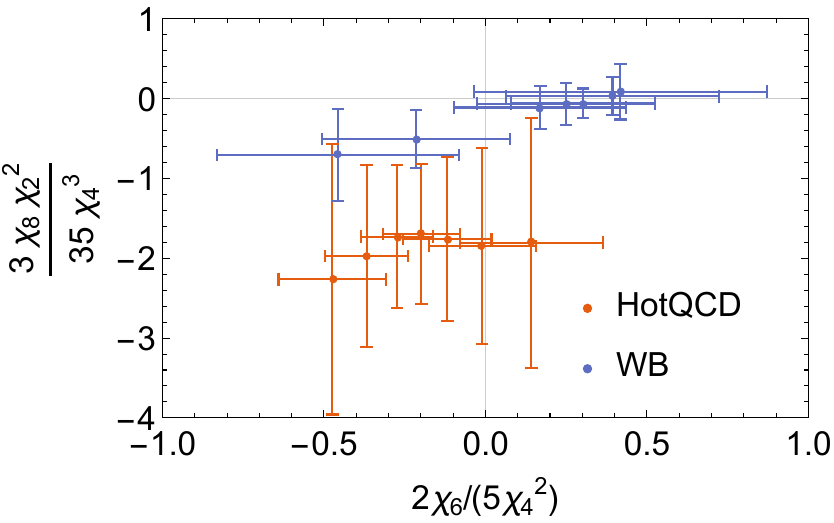}
\caption{The comparison between the Taylor coefficients obtained by the Wuppertal-Budapest (WB) and HotQCD collaborations for the temperature range 135-165 MeV increasing in 5 MeV intervals. For both datasets the temperature increases from right to left.}
\label{fig:cs-WB-compare}
\end{figure}

Even though there are four coefficients in the Taylor expansion, via rescaling $p$ and $\mu_B^2$, one can show that the Pad\'e poles can be expressed as a function of  \textit{two} variables, $\chi_2\chi_6/\chi_4^2$ and $\chi_8\chi_2^2/\chi_4^3$, up to an overall factor of $\chi_2/\chi_4$. By using the same normalization in Ref. \cite{Bollweg:2022rps}, these parameters computed by the two lattice collaborations are shown in Fig. \ref{fig:cs-WB-compare} for the temperature range that this work focuses on. Notably the coefficients are quantitatively different (whereas the overall factor $\chi_2/\chi_4$ varies by a few percent between WB and HotQCD data), signaling the importance of the systematics in any kind of resummation framework that is based on Taylor coefficients. 
   
\begin{figure}[h]
\includegraphics[scale=0.48]{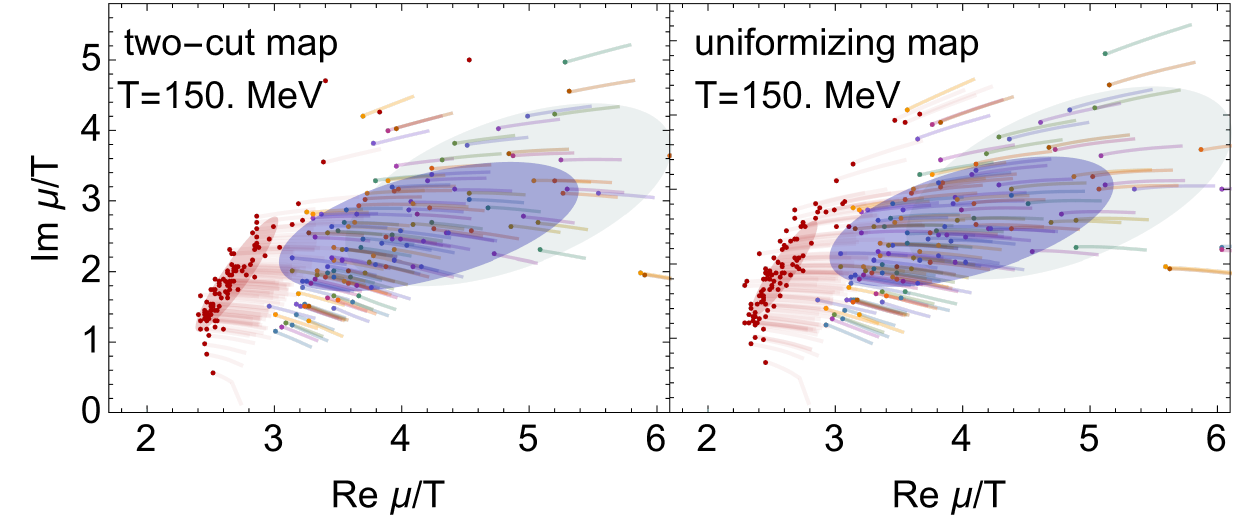}
\caption{Trajectories of the iteration procedure described in Section \ref{sec:results} obtained by using the Wuppertal-Budapest data for  $T=150$ MeV. The blue and green uncertainty regions correspond to Pad\'e and conformal Pad\'e, compared with the results obtained from HotQCD data (in red).}
\label{fig:trajectories-WB-compare}
\end{figure}

Based on the observation that the original Taylor coefficients computed by different collaborations are different, one might expect that the LY singularities obtained from Pad\'e resummations would also differ substantially. However, I found that this is not entirely the case. Before doing so, it is worth commenting on the statistical uncertainties first. Even though the overall statistical errors in the WB results are smaller in magnitude, the overall signal to noise ratio for the LY singularities is actually higher compared to HotQCD results.  This is because $\chi_8$ is closer to zero. As an illustration, the trajectories of the iteration procedure described in Section \ref{sec:results} for two different conformal maps are shown in Fig. \ref{fig:trajectories-WB-compare}. As explained in Section \ref{sec:results}, each line represents a trajectory whose initial point is the Pad\'e singularity associated with the Taylor coefficients sampled from a Gaussian ensemble, and the final point is obtained after 100 iterations of the conformal map.  For comparison, the results obtained from the HotQCD Collaboration are shown in red. The qualitative behavior of the trajectories is similar; however, the WB results have larger statistical uncertainties, most likely due to the small mean value of $\chi_8$ as mentioned.

\begin{figure}[h]
\center
\qquad\includegraphics[scale=0.48]{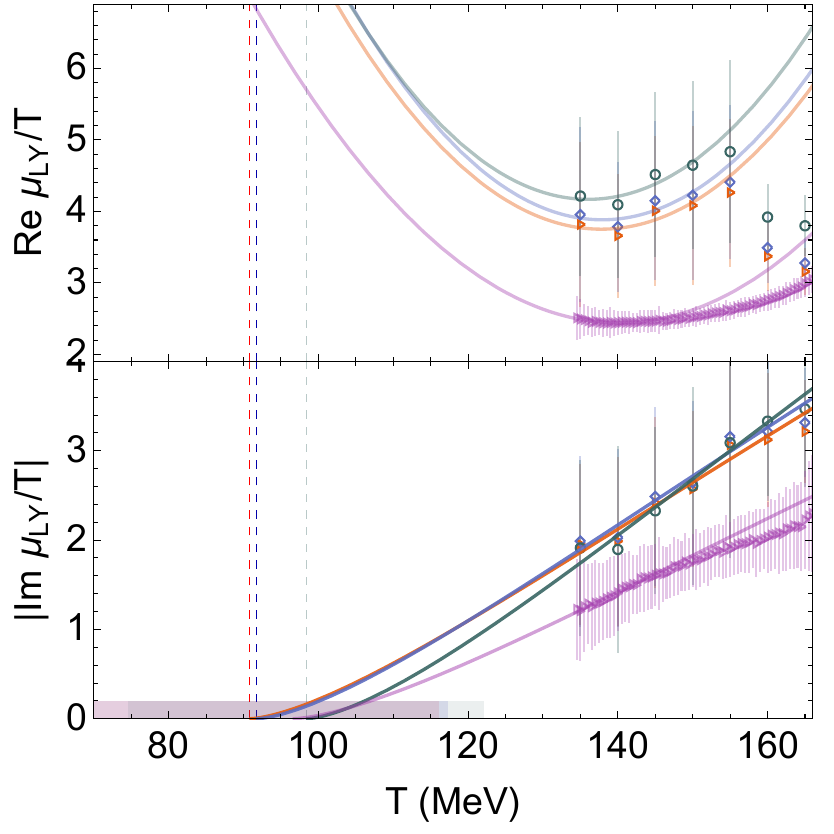}
\includegraphics[scale=0.5]{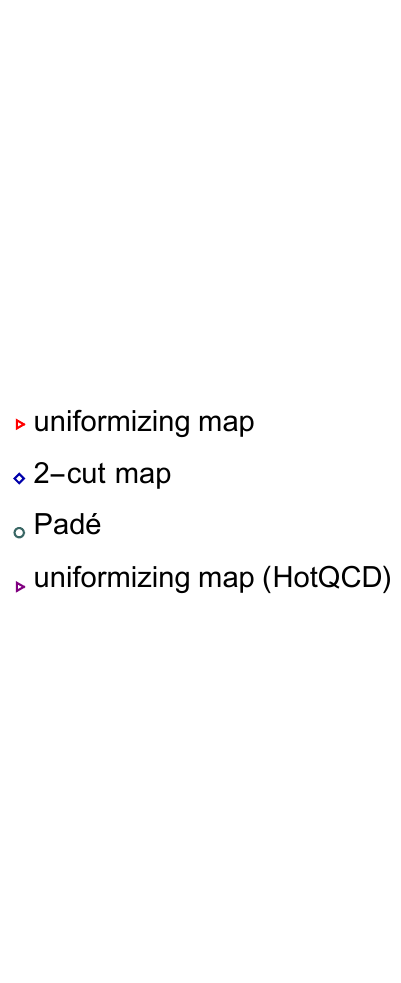}
\caption{The LY trajectory as a function of temperature obtained from the Wuppertal-Budapest data with the same fits given in Eq. \eqref{eq:fits}. For comparison I have included the trajectory obtained from the HotQCD data via the uniformizing map.}
\label{fig:ReImT-WB-compare}
\end{figure}

The LY trajectory for different temperatures is shown in Fig. \ref{fig:ReImT-WB-compare}. Qualitatively, it looks similar to the one obtained from the HotQCD data (see Fig. \ref{fig:LY-traj}). In particular, its imaginary part decreases with decreasing temperature. This behavior is consistent with what one would expect if there is a critical point, since at a critical point the imaginary part vanishes.  Although the error bars are too large to determine the exact functional form of the trajectory with good accuracy, using the functional form assuming the ${\mathbb Z}_2$ scaling form given in Eq. \eqref{eq:ly-traj} leads to reasonable values for the critical temperature listed in Table \ref{table2}.  These results are consistent with those obtained from the HotQCD data (see Table \ref{table}) albeit with larger statistical uncertainties. At the same time, even though the real part of the LY trajectory is qualitatively similar to the one obtained from HotQCD data, the systematic difference is more pronounced. Furthermore, due to the larger statistical uncertainties combined with those that already enter the estimation of $T_c$, it is not possible to meaningfully extract $\mu_c$ with the available data.

\begin{table}
\renewcommand{\arraystretch}{1.7}{\begin{tabular}{  | c  | c | c | c |  c | } 
 \hline
 \, & $T_c$ (MeV) &
  $c_2$   (MeV$^{1-\beta\delta}$) 
 \\
\hline
uniformizing  \,&\,  $91^{+25}_{-25}$  \,&\,  
$2.73^{+0.87}_{-1.81}$ \,
\\  
 two-cut \,&\, $ 92^{+25}_{-25}$   \,&\,  
 $2.87^{+0.92}_{-1.98}$ \,
\\  
Pad\'e \,&\,  $98^{+24}_{-24}$  \,&\,  
$3.43^{+0.99}_{-1.86}$ \,
\\
 \hline
 \end{tabular}
 }
 \caption{$T_c$ and $c_2$ extracted from Pad\'e and conformal Pad\'e using the WB data. The sub/superscripts denote the $1\sigma$ uncertainty. }
 \label{table2}
\end{table}

In conclusion, the most robust feature of the Lee-Yang trajectory constructed from two separate lattice computations, HotQCD and Wuppertal-Budapest, is that its imaginary part decreases temperature, consistent with the existence of a critical point. However, it does not vanish within the available temperature range (within 1$\sigma$ uncertainty); signaling that, if exists, the critical point is likely at $T_c< 130$ MeV. Determining the quantitative features of the LY trajectory is more challenging with the current data, however, the extrapolation of  using the ${\mathbb Z}_2$ scaling ansatz leads to a robust value of $T_c\sim 100$ MeV, both for the WB and HotQCD data. Due to the statistical and systematic uncertainties, estimating the value of the critical chemical potential is even more challenging. At the same time, the current lattice data is consistent with the estimation $\mu_c\sim 600$ MeV.

\section{Summary and Discussion }
\label{sec:discussion}

The task of extracting the location of the Lee-Yang singularities from the presently available lattice QCD data is a challenging one. Currently, there is only access to four coefficients with sizable statistical uncertainties. In this paper, I utilized various conformal maps to improve the accuracy of the usual Pad\'e resummation. By choosing appropriately designed conformal maps, I incorporated further analytical information regarding the equation of state, in addition to the Taylor coefficients; namely the closest singularities must be complex conjugate pair, i.e. the equation of state must be defined on a two-cut Riemann surface. I then conformally mapped this surface into the unit disk and extracted the Pad\'e singularities there. Performing Pad\'e resummation on a compact space gave a better handle in pinning down the location of the true singularity.

 In order to take into account the statistical uncertainties in the Taylor coefficients, they are sampled from a Gaussian ensemble whose variance is matched by the lattice data. The error due to the small number of Taylor coefficients, on the other hand, is harder to quantify. For large number of Taylor coefficients, there is a scaling relation between the magnitude of noise in the Taylor coefficients and the expansion order before (conformal) Pad\'e resummation breaks down \cite{Costin:2022hgc}. However it is an asymptotic result whose applicability to four terms is unclear. 

In order to refine the estimation for the location of the singularity a novel iterative tool is used. Remarkably for both conformal maps, the iteration brought the images of the singularities extracted from conformal Pad\'e closer to the edge of the unit disk where the real singularity lies, as seen in Figs. \ref{fig:iteration-2cut} and  \ref{fig:iteration-uni}. This observation increases the confidence in the results of conformal Pad\'e. Note that the uniformizing map is \textit{exponentially} sensitive to the location of the singularity. Therefore it is not surprising that the poles in $\z$ plane for the uniformizing map are further away from the edge of the unit disk since neither the number Taylor coefficients is large enough nor their precision high enough to resolve singularity with exponential accuracy. It is also worth pointing out that the final results for the LY trajectory obtained from both conformal maps agree with each other and systematically differ from those from ordinary Pad\'e, with the conformal Pad\'e results being slightly less sensitive to the statistical uncertainties compared to Pad\'e which can be seen in Figs. \ref{fig:iteration-2cut}, \ref{fig:iteration-uni} and \ref{eq:ly-traj}. Furthermore, as explained in e.g. Ref. \cite{Costin:2020pcj} conformal maps provide a better estimate for the singularity compared to ordinary Pad\'e. The various test cases where the ordinary and conformal  Pad\'e predictions can be compared with the exact results also confirm this observation \cite{Basar:2021hdf, Basar:2021gyi}. For these reasons, I think that conformal Pad\'e results are more trustworthy than those of Pad\'e.

Then, from the LY trajectory constructed via the resummations, I extrapolated the location of the critical point, as well as constrained values of the non-universal mapping parameters in its vicinity. These results are given in Table \ref{table}. This is the central result of this paper.  Notice that  several significant figures are used in these extrapolations. This is not to claim such precision in the final estimate, but to illustrate the quantitative differences between different resummations. The fact that they all lie in the same overall region is encouraging. Furthermore, the results obtained from two independent lattice computations indicate a robust trend in $\im \mu_L{LY}$ consistent with $T_c\approx 100$ MeV.  My results also agree with other results computed via similar Pad\'e type resummations \cite{Clarke:2023noy}, as well as other methods such as functional renormalization group, ($T_c , \mu_C$) = (107, 635) MeV \cite{Fu:2019hdw} and truncated Dyson Schwinger equations, ($T_c , \mu_C$) = (117, 600) MeV \cite{Bernhardt:2021iql,Fischer21}. The estimate based on the best fit, $\mu_c/T_c\approx 6$, is also consistent with the constraints from the recent lattice QCD data which strongly disfavor the existence of a critical point for $\mu_c/T_c\lesssim 3$ \cite{Borsanyi:2020fev}. 

Because the extrapolated $T_c$ ($\sim$ 100 MeV) is fairly lower than the minimum available temperature (135 MeV) I had to rely on best fits for the estimates for the critical point. Especially  $\re\mu_{LY}$, from which  $\mu_c$ is extrapolated, depends sensitively on the quadratic fit. Combined with the current statistical and systematic uncertainties in the Taylor coefficients, such an extrapolation introduces a sizable uncertainty in the estimate for $\mu_c$. Having access to lattice data for lower temperatures in the future would reduce the sensitive reliance on the extrapolation and therefore significantly improve the accuracy of the estimation of the location of the critical point.

Another important point worth discussing is that I assumed the equation of state obeys the scaling obtained from the universality class of the ``$Z_2$ critical point" for part of the available temperature range for the LY trajectory in order to extrapolate the data to $\mu_c$ and  $T_c$. In this scaling, the relevant axes mapped to the Ising parameters $h$ and $r$, are identified as $\mu_B$ and $T$. At the moment, it is unclear whether this scaling is valid at these temperatures. It is widely believed that around the pseudo-critical temperature, the closest singularity is associated with the ``$O(4)$ critical point" (more precisely a line of second order points) that is located at $m_{u,d}=0$ and belongs to the $O(4)$ universality class.  Therefore the LY trajectory should obey the ``$O(4)$ scaling''  which has a different form than the $Z_2$ scaling used in Eq. \eqref{eq:ly-traj}. This is because the $O(4)$ scaling identifies one of relevant axes with the quark mass. Even though the agreement between the data and the form predicted by the $Z_2$ scaling seems suggestive that, at least for $T\in(130-145)$ MeV range, the $Z_2$ scaling is valid, this point requires further investigation. This issue will be addressed in a forthcoming publication. 

Finally, a related issue concerns the shape of the LY trajectory. In various models that exhibit similar critical phenomena to the conjectured QCD phase diagram, the LY trajectory can be exactly calculated. Among them are the Gross-Neveu model  \cite{Basar:2021hdf}, quark meson model \cite{Mukherjee:2021tyg} and the Chiral Random Matrix model \cite{Stephanov:2006dn, Basar:2021gyi} where $\re \mu_{LY}(T)$ and $\im \mu_{LY}(T)$ are monotonically decreasing and increasing functions of $T$ respectively, for $T>T_c$. As seen in Fig. \ref{fig:LY-traj}, the QCD data indicates that $\re \mu_{LY}$ is \textit{not} monotonically decreasing in $T$.  For $T\gtrsim 145$ MeV, it is actually \textit{increasing}. If this behavior is indeed representative of the physical LY trajectory and not an artifact of the resummations due to a combination of noise and small number of Taylor coefficients, it asks for further investigation. In fact, at higher temperatures, the shape of the LY trajectory is expected to be controlled by yet another singular point; the Roberge-Weiss singularity \cite{Roberge:1986mm} and for some $T=T_{RW}$, the trajectory should pass through the point(s) $\re \mu_{LY}(T_{RW})=0$, $\im \mu_{LY}(T_{RW})=\pm \pi$.  A recent estimate for the Roberge-Weiss temperature based on multi-point Pad\'e approximants is $T_{RW}\approx 170$ MeV \cite{Schmidt:2022ogw}.  This means that, if the non-monotonic behavior is correct, then for some  $T\in  (160- 170)$ MeV , $\re \mu_{LY}(T)$ has to peak and then go down to zero. At the same time, because $\chi_8$ crosses zero around $T\approx170$ MeV, the Pad\'e singularities are too noisy around these temperatures to lead to any meaningful estimation of $\mu_{LY}$. Note that the Roberge-Weiss point is related to confinement/deconfinement transition and therefore is not present in any of the aforementioned models. This discussion is left for future work as well.

\section{Acknowledgments}
I thank M. Stephanov and G. Dunne for fruitful discussions. The author is supported by the National Science Foundation CAREER Award No. PHY-2143149.

\bibliographystyle{utphys}
\bibliography{references}

\providecommand{\href}[2]{#2}\begingroup\raggedright\begin{thebibliography}{10}

\bibitem{Aoki:2006we}
Y.~Aoki, G.~Endrodi, Z.~Fodor, S.~D. Katz, and K.~K. Szabo, ``{The Order of the
  quantum chromodynamics transition predicted by the standard model of particle
  physics},'' \href{http://dx.doi.org/10.1038/nature05120}{{\em Nature}
  {\bfseries 443} (2006) 675--678},
  \href{http://arxiv.org/abs/hep-lat/0611014}{{\ttfamily
  arXiv:hep-lat/0611014}}.

\bibitem{Philipsen:2007aa}
O.~Philipsen, ``Lattice qcd at finite temperature and density,''
  \href{http://dx.doi.org/10.1140/epjst/e2007-00376-3}{{\em The European
  Physical Journal Special Topics} {\bfseries 152} no.~1, (2007) 29--60}.
  \url{https://doi.org/10.1140/epjst/e2007-00376-3}.

\bibitem{deForcrand:2009zkb}
P.~de~Forcrand, ``{Simulating QCD at finite density},''
  \href{http://dx.doi.org/10.22323/1.091.0010}{{\em PoS} {\bfseries LAT2009}
  (2009) 010}, \href{http://arxiv.org/abs/1005.0539}{{\ttfamily arXiv:1005.0539
  [hep-lat]}}.

\bibitem{Ding:2015ona}
H.-T. Ding, F.~Karsch, and S.~Mukherjee, ``{Thermodynamics of
  strong-interaction matter from Lattice QCD},''
  \href{http://dx.doi.org/10.1142/S0218301315300076}{{\em Int. J. Mod. Phys. E}
  {\bfseries 24} no.~10, (2015) 1530007},
  \href{http://arxiv.org/abs/1504.05274}{{\ttfamily arXiv:1504.05274
  [hep-lat]}}.

\bibitem{Bzdak:2019pkr}
A.~Bzdak, S.~Esumi, V.~Koch, J.~Liao, M.~Stephanov, and N.~Xu, ``{Mapping the
  Phases of Quantum Chromodynamics with Beam Energy Scan},''
  \href{http://dx.doi.org/10.1016/j.physrep.2020.01.005}{{\em Phys. Rept.}
  {\bfseries 853} (2020) 1--87},
  \href{http://arxiv.org/abs/1906.00936}{{\ttfamily arXiv:1906.00936
  [nucl-th]}}.

\bibitem{Almaalol:2022xwv}
D.~Almaalol {\em et~al.}, ``{QCD Phase Structure and Interactions at High
  Baryon Density: Continuation of BES Physics Program with CBM at FAIR},''
  \href{http://arxiv.org/abs/2209.05009}{{\ttfamily arXiv:2209.05009
  [nucl-ex]}}.

\bibitem{Allton:2002zi}
C.~R. Allton, S.~Ejiri, S.~J. Hands, O.~Kaczmarek, F.~Karsch, E.~Laermann,
  C.~Schmidt, and L.~Scorzato, ``{The QCD thermal phase transition in the
  presence of a small chemical potential},''
  \href{http://dx.doi.org/10.1103/PhysRevD.66.074507}{{\em Phys. Rev. D}
  {\bfseries 66} (2002) 074507},
  \href{http://arxiv.org/abs/hep-lat/0204010}{{\ttfamily
  arXiv:hep-lat/0204010}}.

\bibitem{deForcrand:2002hgr}
P.~de~Forcrand and O.~Philipsen, ``{The QCD phase diagram for small densities
  from imaginary chemical potential},''
  \href{http://dx.doi.org/10.1016/S0550-3213(02)00626-0}{{\em Nucl. Phys. B}
  {\bfseries 642} (2002) 290--306},
  \href{http://arxiv.org/abs/hep-lat/0205016}{{\ttfamily
  arXiv:hep-lat/0205016}}.

\bibitem{DElia:2002tig}
M.~D'Elia and M.-P. Lombardo, ``{Finite density QCD via imaginary chemical
  potential},'' \href{http://dx.doi.org/10.1103/PhysRevD.67.014505}{{\em Phys.
  Rev. D} {\bfseries 67} (2003) 014505},
  \href{http://arxiv.org/abs/hep-lat/0209146}{{\ttfamily
  arXiv:hep-lat/0209146}}.

\bibitem{Bellwied:2015rza}
R.~Bellwied, S.~Borsanyi, Z.~Fodor, J.~G\"unther, S.~D. Katz, C.~Ratti, and
  K.~K. Szabo, ``{The QCD phase diagram from analytic continuation},''
  \href{http://dx.doi.org/10.1016/j.physletb.2015.11.011}{{\em Phys. Lett. B}
  {\bfseries 751} (2015) 559--564},
  \href{http://arxiv.org/abs/1507.07510}{{\ttfamily arXiv:1507.07510
  [hep-lat]}}.

\bibitem{Ratti:2018ksb}
C.~Ratti, ``{Lattice QCD and heavy ion collisions: a review of recent
  progress},'' \href{http://dx.doi.org/10.1088/1361-6633/aabb97}{{\em Rept.
  Prog. Phys.} {\bfseries 81} no.~8, (2018) 084301},
  \href{http://arxiv.org/abs/1804.07810}{{\ttfamily arXiv:1804.07810
  [hep-lat]}}.

\bibitem{Borsanyi:2021sxv}
S.~Bors\'anyi, Z.~Fodor, J.~N. Guenther, R.~Kara, S.~D. Katz, P.~Parotto,
  A.~P\'asztor, C.~Ratti, and K.~K. Szab\'o, ``{Lattice QCD equation of state
  at finite chemical potential from an alternative expansion scheme},''
  \href{http://arxiv.org/abs/2102.06660}{{\ttfamily arXiv:2102.06660
  [hep-lat]}}.

\bibitem{Bollweg:2022rps}
{\bfseries HotQCD} Collaboration, D.~Bollweg, J.~Goswami, O.~Kaczmarek,
  F.~Karsch, S.~Mukherjee, P.~Petreczky, C.~Schmidt, and P.~Scior, ``{Taylor
  expansions and Pad\'e approximants for cumulants of conserved charge
  fluctuations at nonvanishing chemical potentials},''
  \href{http://dx.doi.org/10.1103/PhysRevD.105.074511}{{\em Phys. Rev. D}
  {\bfseries 105} no.~7, (2022) 074511},
  \href{http://arxiv.org/abs/2202.09184}{{\ttfamily arXiv:2202.09184
  [hep-lat]}}.

\bibitem{Bollweg:2022fqq}
{\bfseries HotQCD} Collaboration, D.~Bollweg, D.~A. Clarke, J.~Goswami,
  O.~Kaczmarek, F.~Karsch, S.~Mukherjee, P.~Petreczky, C.~Schmidt, and
  S.~Sharma, ``{Equation of state and speed of sound of (2+1)-flavor QCD in
  strangeness-neutral matter at nonvanishing net baryon-number density},''
  \href{http://dx.doi.org/10.1103/PhysRevD.108.014510}{{\em Phys. Rev. D}
  {\bfseries 108} no.~1, (2023) 014510},
  \href{http://arxiv.org/abs/2212.09043}{{\ttfamily arXiv:2212.09043
  [hep-lat]}}.

\bibitem{Borsanyi:2020fev}
S.~Borsanyi, Z.~Fodor, J.~N. Guenther, R.~Kara, S.~D. Katz, P.~Parotto,
  A.~Pasztor, C.~Ratti, and K.~K. Szabo, ``{QCD Crossover at Finite Chemical
  Potential from Lattice Simulations},''
  \href{http://dx.doi.org/10.1103/PhysRevLett.125.052001}{{\em Phys. Rev.
  Lett.} {\bfseries 125} no.~5, (2020) 052001},
  \href{http://arxiv.org/abs/2002.02821}{{\ttfamily arXiv:2002.02821
  [hep-lat]}}.

\bibitem{Fisher:1974series}
M.~E. Fisher, ``Critical point phenomena - the role of series expansions,''
  {\em Rocky Mountain Journal of Mathematics} {\bfseries 4} no.~2, (1974) 181.

\bibitem{Halasz_1997}
M.~Halasz, A.~Jackson, and J.~Verbaarschot, ``Yang-lee zeros of a random matrix
  model for qcd at finite density,''
  \href{http://dx.doi.org/10.1016/s0370-2693(97)00015-4}{{\em Physics Letters
  B} {\bfseries 395} no.~3-4, (Mar, 1997) 293--297}.
  \url{http://dx.doi.org/10.1016/S0370-2693(97)00015-4}.

\bibitem{Ejiri:2005ts}
S.~Ejiri, ``{Lee-Yang zero analysis for the study of QCD phase structure},''
  \href{http://dx.doi.org/10.1103/PhysRevD.73.054502}{{\em Phys. Rev. D}
  {\bfseries 73} (2006) 054502},
  \href{http://arxiv.org/abs/hep-lat/0506023}{{\ttfamily
  arXiv:hep-lat/0506023}}.

\bibitem{Stephanov:2006dn}
M.~A. Stephanov, ``{QCD critical point and complex chemical potential
  singularities},'' \href{http://dx.doi.org/10.1103/PhysRevD.73.094508}{{\em
  Phys. Rev. D} {\bfseries 73} (2006) 094508},
  \href{http://arxiv.org/abs/hep-lat/0603014}{{\ttfamily
  arXiv:hep-lat/0603014}}.

\bibitem{Mukherjee:2019eou}
S.~Mukherjee and V.~Skokov, ``{Universality driven analytic structure of the
  QCD crossover: radius of convergence in the baryon chemical potential},''
  \href{http://dx.doi.org/10.1103/PhysRevD.103.L071501}{{\em Phys. Rev. D}
  {\bfseries 103} no.~7, (2021) L071501},
  \href{http://arxiv.org/abs/1909.04639}{{\ttfamily arXiv:1909.04639
  [hep-ph]}}.

\bibitem{Connelly:2020pno}
A.~Connelly, G.~Johnson, S.~Mukherjee, and V.~Skokov, ``{Universality driven
  analytic structure of QCD crossover: radius of convergence and QCD critical
  point},'' \href{http://dx.doi.org/10.1016/j.nuclphysa.2020.121834}{{\em Nucl.
  Phys. A} {\bfseries 1005} (2021) 121834},
  \href{http://arxiv.org/abs/2004.05095}{{\ttfamily arXiv:2004.05095
  [hep-ph]}}.

\bibitem{Basar:2021hdf}
G.~Basar, ``{Universality, Lee-Yang Singularities, and Series Expansions},''
  \href{http://dx.doi.org/10.1103/PhysRevLett.127.171603}{{\em Phys. Rev.
  Lett.} {\bfseries 127} no.~17, (2021) 171603},
  \href{http://arxiv.org/abs/2105.08080}{{\ttfamily arXiv:2105.08080
  [hep-th]}}.

\bibitem{Basar:2021gyi}
G.~Basar, G.~V. Dunne, and Z.~Yin, ``{Uniformizing Lee-Yang singularities},''
  \href{http://dx.doi.org/10.1103/PhysRevD.105.105002}{{\em Phys. Rev. D}
  {\bfseries 105} no.~10, (2022) 105002},
  \href{http://arxiv.org/abs/2112.14269}{{\ttfamily arXiv:2112.14269
  [hep-th]}}.

\bibitem{Dimopoulos:2021vrk}
P.~Dimopoulos, L.~Dini, F.~Di~Renzo, J.~Goswami, G.~Nicotra, C.~Schmidt,
  S.~Singh, K.~Zambello, and F.~Ziesch\'e, ``{Contribution to understanding the
  phase structure of strong interaction matter: Lee-Yang edge singularities
  from lattice QCD},''
  \href{http://dx.doi.org/10.1103/PhysRevD.105.034513}{{\em Phys. Rev. D}
  {\bfseries 105} no.~3, (2022) 034513},
  \href{http://arxiv.org/abs/2110.15933}{{\ttfamily arXiv:2110.15933
  [hep-lat]}}.

\bibitem{Nicotra:2021ijp}
G.~Nicotra, P.~Dimopoulos, L.~Dini, F.~Di~Renzo, J.~Goswami, C.~Schmidt,
  S.~Singh, K.~Zambello, and F.~Ziesche, ``{Lee-Yang edge singularities in 2+1
  flavor QCD with imaginary chemical potential.},''
  \href{http://dx.doi.org/10.22323/1.396.0260}{{\em PoS} {\bfseries
  LATTICE2021} (2022) 260}, \href{http://arxiv.org/abs/2111.05630}{{\ttfamily
  arXiv:2111.05630 [hep-lat]}}.

\bibitem{Singh:2021pog}
{\bfseries Bielefeld-Parma} Collaboration, S.~Singh, P.~Dimopoulos, L.~Dini,
  F.~Di~Renzo, J.~Goswami, G.~Nicotra, C.~Schmidt, K.~Zambello, and F.~Ziesche,
  ``Lee-yang edge singularities in lattice qcd : A systematic study of
  singularities in the complex mub plane using rational approximations.,''
  \href{http://dx.doi.org/10.22323/1.396.0544}{{\em PoS} {\bfseries
  LATTICE2021} (2022) 544}, \href{http://arxiv.org/abs/2111.06241}{{\ttfamily
  arXiv:2111.06241 [hep-lat]}}.

\bibitem{Schmidt:2022ogw}
C.~Schmidt, D.~A. Clarke, G.~Nicotra, F.~Di~Renzo, P.~Dimopoulos, S.~Singh,
  J.~Goswami, and K.~Zambello, ``{Detecting Critical Points from the
  Lee\textendash{}Yang Edge Singularities in Lattice QCD},''
  \href{http://dx.doi.org/10.5506/APhysPolBSupp.16.1-A52}{{\em Acta Phys.
  Polon. Supp.} {\bfseries 16} no.~1, (2023) 1--A52},
  \href{http://arxiv.org/abs/2209.04345}{{\ttfamily arXiv:2209.04345
  [hep-lat]}}.

\bibitem{Clarke:2023noy}
D.~A. Clarke, K.~Zambello, P.~Dimopoulos, F.~Di~Renzo, J.~Goswami, G.~Nicotra,
  C.~Schmidt, and S.~Singh, ``{Determination of Lee-Yang edge singularities in
  QCD by rational approximations},''
  \href{http://dx.doi.org/10.22323/1.430.0164}{{\em PoS} {\bfseries
  LATTICE2022} (2023) 164}, \href{http://arxiv.org/abs/2301.03952}{{\ttfamily
  arXiv:2301.03952 [hep-lat]}}.

\bibitem{Parotto:2018pwx}
P.~Parotto, M.~Bluhm, D.~Mroczek, M.~Nahrgang, J.~Noronha-Hostler,
  K.~Rajagopal, C.~Ratti, T.~Sch\"afer, and M.~Stephanov, ``{QCD equation of
  state matched to lattice data and exhibiting a critical point singularity},''
  \href{http://dx.doi.org/10.1103/PhysRevC.101.034901}{{\em Phys. Rev. C}
  {\bfseries 101} no.~3, (2020) 034901},
  \href{http://arxiv.org/abs/1805.05249}{{\ttfamily arXiv:1805.05249
  [hep-ph]}}.

\bibitem{An:2021wof}
X.~An {\em et~al.}, ``{The BEST framework for the search for the QCD critical
  point and the chiral magnetic effect},''
  \href{http://dx.doi.org/10.1016/j.nuclphysa.2021.122343}{{\em Nucl. Phys. A}
  {\bfseries 1017} (2022) 122343},
  \href{http://arxiv.org/abs/2108.13867}{{\ttfamily arXiv:2108.13867
  [nucl-th]}}.

\bibitem{Borsanyi:2018grb}
S.~Borsanyi, Z.~Fodor, J.~N. Guenther, S.~K. Katz, K.~K. Szabo, A.~Pasztor,
  I.~Portillo, and C.~Ratti, ``{Higher order fluctuations and correlations of
  conserved charges from lattice QCD},''
  \href{http://dx.doi.org/10.1007/JHEP10(2018)205}{{\em JHEP} {\bfseries 10}
  (2018) 205}, \href{http://arxiv.org/abs/1805.04445}{{\ttfamily
  arXiv:1805.04445 [hep-lat]}}.

\bibitem{Yang:1952be}
C.-N. Yang and T.~D. Lee, ``{Statistical theory of equations of state and phase
  transitions. 1. Theory of condensation},''
  \href{http://dx.doi.org/10.1103/PhysRev.87.404}{{\em Phys. Rev.} {\bfseries
  87} (1952) 404--409}.

\bibitem{Lee:1952ig}
T.~D. Lee and C.-N. Yang, ``{Statistical theory of equations of state and phase
  transitions. 2. Lattice gas and Ising model},''
  \href{http://dx.doi.org/10.1103/PhysRev.87.410}{{\em Phys. Rev.} {\bfseries
  87} (1952) 410--419}.

\bibitem{ZinnJustin:2002ru}
J.~Zinn-Justin, ``{Quantum field theory and critical phenomena},'' {\em Int.
  Ser. Monogr. Phys.} {\bfseries 113} (2002) 1--1054.

\bibitem{PhysRevD.86.025022}
S.~El-Showk, M.~F. Paulos, D.~Poland, S.~Rychkov, D.~Simmons-Duffin, and
  A.~Vichi, ``Solving the 3d ising model with the conformal bootstrap,''
  \href{http://dx.doi.org/10.1103/PhysRevD.86.025022}{{\em Phys. Rev. D}
  {\bfseries 86} (Jul, 2012) 025022}.
  \url{https://link.aps.org/doi/10.1103/PhysRevD.86.025022}.

\bibitem{Rennecke:2022ohx}
F.~Rennecke and V.~V. Skokov, ``{Universal location of Yang\textendash{}Lee
  edge singularity for a one-component field theory in
  1\ensuremath{\leq}d\ensuremath{\leq}4},''
  \href{http://dx.doi.org/10.1016/j.aop.2022.169010}{{\em Annals Phys.}
  {\bfseries 444} (2022) 169010},
  \href{http://arxiv.org/abs/2203.16651}{{\ttfamily arXiv:2203.16651
  [hep-ph]}}.

\bibitem{Connelly:2020gwa}
A.~Connelly, G.~Johnson, F.~Rennecke, and V.~Skokov, ``{Universal Location of
  the Yang-Lee Edge Singularity in $O(N)$ Theories},''
  \href{http://dx.doi.org/10.1103/PhysRevLett.125.191602}{{\em Phys. Rev.
  Lett.} {\bfseries 125} no.~19, (2020) 191602},
  \href{http://arxiv.org/abs/2006.12541}{{\ttfamily arXiv:2006.12541
  [cond-mat.stat-mech]}}.

\bibitem{Johnson:2022cqv}
G.~Johnson, F.~Rennecke, and V.~V. Skokov, ``{Universal location of Yang-Lee
  edge singularity in classic O(N) universality classes},''
  \href{http://dx.doi.org/10.1103/PhysRevD.107.116013}{{\em Phys. Rev. D}
  {\bfseries 107} no.~11, (2023) 116013},
  \href{http://arxiv.org/abs/2211.00710}{{\ttfamily arXiv:2211.00710
  [hep-ph]}}.

\bibitem{Karsch:2023rfb}
F.~Karsch, C.~Schmidt, and S.~Singh, ``{Lee-Yang and Langer edge singularities
  from analytic continuation of scaling functions},''
  \href{http://arxiv.org/abs/2311.13530}{{\ttfamily arXiv:2311.13530
  [hep-lat]}}.

\bibitem{Fisher:1978pf}
M.~E. Fisher, ``{Yang-Lee Edge Singularity and phi**3 Field Theory},''
  \href{http://dx.doi.org/10.1103/PhysRevLett.40.1610}{{\em Phys. Rev. Lett.}
  {\bfseries 40} (1978) 1610--1613}.

\bibitem{An:2016lni}
X.~An, D.~Mesterh\'azy, and M.~A. Stephanov, ``{Functional renormalization
  group approach to the Yang-Lee edge singularity},''
  \href{http://dx.doi.org/10.1007/JHEP07(2016)041}{{\em JHEP} {\bfseries 07}
  (2016) 041}, \href{http://arxiv.org/abs/1605.06039}{{\ttfamily
  arXiv:1605.06039 [hep-th]}}.

\bibitem{An:2017brc}
X.~An, D.~Mesterh\'azy, and M.~A. Stephanov, ``{On spinodal points and Lee-Yang
  edge singularities},'' \href{http://dx.doi.org/10.1088/1742-5468/aaac4a}{{\em
  J. Stat. Mech.} {\bfseries 1803} no.~3, (2018) 033207},
  \href{http://arxiv.org/abs/1707.06447}{{\ttfamily arXiv:1707.06447
  [hep-th]}}.

\bibitem{Fonseca:2001dc}
P.~Fonseca and A.~Zamolodchikov, ``{Ising field theory in a magnetic field:
  Analytic properties of the free energy},''
  \href{http://arxiv.org/abs/hep-th/0112167}{{\ttfamily arXiv:hep-th/0112167}}.

\bibitem{Pradeep:2019ccv}
M.~S. Pradeep and M.~Stephanov, ``{Universality of the critical point mapping
  between Ising model and QCD at small quark mass},''
  \href{http://dx.doi.org/10.1103/PhysRevD.100.056003}{{\em Phys. Rev. D}
  {\bfseries 100} no.~5, (2019) 056003},
  \href{http://arxiv.org/abs/1905.13247}{{\ttfamily arXiv:1905.13247
  [hep-ph]}}.

\bibitem{Bazavov:2017dus}
A.~Bazavov {\em et~al.}, ``{The QCD Equation of State to $\mathcal{O}(\mu_B^6)$
  from Lattice QCD},'' \href{http://dx.doi.org/10.1103/PhysRevD.95.054504}{{\em
  Phys. Rev. D} {\bfseries 95} no.~5, (2017) 054504},
  \href{http://arxiv.org/abs/1701.04325}{{\ttfamily arXiv:1701.04325
  [hep-lat]}}.

\bibitem{GIORDANO2021121986}
M.~Giordano, K.~Kapas, S.~Katz, D.~Nogradi, and A.~Pasztor, ``Towards a
  reliable lower bound on the location of the critical endpoint,''
  \href{http://dx.doi.org/https://doi.org/10.1016/j.nuclphysa.2020.121986}{{\em
  Nuclear Physics A} {\bfseries 1005} (2021) 121986}.
  \url{https://www.sciencedirect.com/science/article/pii/S0375947420302967}.
  The 28th International Conference on Ultra-relativistic Nucleus-Nucleus
  Collisions: Quark Matter 2019.

\bibitem{STAHL1997139}
H.~Stahl, ``The convergence of pad{\'e} approximants to functions with branch
  points,''
  \href{http://dx.doi.org/https://doi.org/10.1006/jath.1997.3141}{{\em Journal
  of Approximation Theory} {\bfseries 91} no.~2, (1997) 139--204}.
  \url{https://www.sciencedirect.com/science/article/pii/S0021904597931415}.

\bibitem{saff}
E.~Saff {\em Surveys in Approximation Theory} {\bfseries 5} (2010) 165--200.

\bibitem{Costin:2020pcj}
O.~Costin and G.~V. Dunne, ``{Uniformization and Constructive Analytic
  Continuation of Taylor Series},''
  \href{http://arxiv.org/abs/2009.01962}{{\ttfamily arXiv:2009.01962
  [math.CV]}}.

\bibitem{Costin:2020hwg}
O.~Costin and G.~V. Dunne, ``{Physical Resurgent Extrapolation},''
  \href{http://dx.doi.org/10.1016/j.physletb.2020.135627}{{\em Phys. Lett. B}
  {\bfseries 808} (2020) 135627},
  \href{http://arxiv.org/abs/2003.07451}{{\ttfamily arXiv:2003.07451
  [hep-th]}}.

\bibitem{Costin:2021bay}
O.~Costin and G.~V. Dunne, ``{Conformal and uniformizing maps in Borel
  analysis},'' \href{http://dx.doi.org/10.1140/epjs/s11734-021-00267-x}{{\em
  Eur. Phys. J. ST} {\bfseries 230} no.~12-13, (2021) 2679--2690},
  \href{http://arxiv.org/abs/2108.01145}{{\ttfamily arXiv:2108.01145
  [hep-th]}}.

\bibitem{Guida:1998bx}
R.~Guida and J.~Zinn-Justin, ``{Critical exponents of the N vector model},''
  \href{http://dx.doi.org/10.1088/0305-4470/31/40/006}{{\em J. Phys. A}
  {\bfseries 31} (1998) 8103--8121},
  \href{http://arxiv.org/abs/cond-mat/9803240}{{\ttfamily
  arXiv:cond-mat/9803240}}.

\bibitem{Rossi:2018}
R.~Rossi, T.~Ohgoe, K.~Van~Houcke, and F.~Werner, ``Resummation of diagrammatic
  series with zero convergence radius for strongly correlated fermions,''
  \href{http://dx.doi.org/10.1103/PhysRevLett.121.130405}{{\em Phys. Rev.
  Lett.} {\bfseries 121} (Sep, 2018) 130405}.
  \url{https://link.aps.org/doi/10.1103/PhysRevLett.121.130405}.

\bibitem{Serone:2019szm}
M.~Serone, G.~Spada, and G.~Villadoro, ``{$\lambda \phi_2^4$ theory
  \textemdash{} Part II. the broken phase beyond NNNN(NNNN)LO},''
  \href{http://dx.doi.org/10.1007/JHEP05(2019)047}{{\em JHEP} {\bfseries 05}
  (2019) 047}, \href{http://arxiv.org/abs/1901.05023}{{\ttfamily
  arXiv:1901.05023 [hep-th]}}.

\bibitem{Costin:2022hgc}
O.~Costin, G.~V. Dunne, and M.~Meynig, ``{Noise effects on Pad\'e approximants
  and conformal maps $^{*}$},''
  \href{http://dx.doi.org/10.1088/1751-8121/aca303}{{\em J. Phys. A} {\bfseries
  55} no.~46, (2022) 464007}, \href{http://arxiv.org/abs/2208.02410}{{\ttfamily
  arXiv:2208.02410 [math-ph]}}.

\bibitem{Fu:2019hdw}
W.-j. Fu, J.~M. Pawlowski, and F.~Rennecke, ``{QCD phase structure at finite
  temperature and density},''
  \href{http://dx.doi.org/10.1103/PhysRevD.101.054032}{{\em Phys. Rev. D}
  {\bfseries 101} no.~5, (2020) 054032},
  \href{http://arxiv.org/abs/1909.02991}{{\ttfamily arXiv:1909.02991
  [hep-ph]}}.

\bibitem{Bernhardt:2021iql}
J.~Bernhardt, C.~S. Fischer, P.~Isserstedt, and B.-J. Schaefer, ``{Critical
  endpoint of QCD in a finite volume},''
  \href{http://dx.doi.org/10.1103/PhysRevD.104.074035}{{\em Phys. Rev. D}
  {\bfseries 104} no.~7, (2021) 074035},
  \href{http://arxiv.org/abs/2107.05504}{{\ttfamily arXiv:2107.05504
  [hep-ph]}}.

\bibitem{Fischer21}
P.~J. Gunkel and C.~S. Fischer, ``Locating the critical endpoint of qcd:
  Mesonic backcoupling effects,''
  \href{http://dx.doi.org/10.1103/PhysRevD.104.054022}{{\em Phys. Rev. D}
  {\bfseries 104} (Sep, 2021) 054022}.
  \url{https://link.aps.org/doi/10.1103/PhysRevD.104.054022}.

\bibitem{Mukherjee:2021tyg}
S.~Mukherjee, F.~Rennecke, and V.~V. Skokov, ``{Analytical structure of the
  equation of state at finite density: Resummation versus expansion in a low
  energy model},'' \href{http://dx.doi.org/10.1103/PhysRevD.105.014026}{{\em
  Phys. Rev. D} {\bfseries 105} no.~1, (2022) 014026},
  \href{http://arxiv.org/abs/2110.02241}{{\ttfamily arXiv:2110.02241
  [hep-ph]}}.

\bibitem{Roberge:1986mm}
A.~Roberge and N.~Weiss, ``{Gauge Theories With Imaginary Chemical Potential
  and the Phases of {QCD}},''
  \href{http://dx.doi.org/10.1016/0550-3213(86)90582-1}{{\em Nucl. Phys. B}
  {\bfseries 275} (1986) 734--745}.

\end{thebibliography}\endgroup

\end{document}